\documentclass[11pt,a4paper]{article}
\pdfoutput=1

\usepackage{jcappub}
\usepackage[utf8]{inputenc}
\usepackage{graphicx}
\usepackage{epsfig}
\usepackage{subfigure}
\usepackage{color}
\usepackage{comment}

\def\beq{\begin{equation}}
\def\eeq{\end{equation}}
\def\baq{\begin{eqnarray}}
\def\eaq{\end{eqnarray}}

\newcommand{\be}{\begin{equation}} 
\newcommand{\ee}{\end{equation}}
\newcommand{\bea}{\begin{equation}\begin{aligned}} 
\newcommand{\eea}{\end{aligned}\end{equation}}

\newcommand{\bmp}{\noindent\begin{minipage}{16cm}}
\newcommand{\emp}{\end{minipage}\vskip 7mm} 
\def\lsim{\mathrel{\raise.3ex\hbox{$<$\kern-.75em\lower1ex\hbox{$\sim$}}}}
\def\gsim{\mathrel{\raise.3ex\hbox{$>$\kern-.75em\lower1ex\hbox{$\sim$}}}}

\newcommand{\intron}[1]{}

\title{The Dawn of FIMP Dark Matter:\\A Review of Models and Constraints}

\author[a, b]{Nicol\'{a}s Bernal,}
\author[c]{Matti Heikinheimo,}
\author[d]{Tommi Tenkanen,}
\author[c]{\\Kimmo Tuominen}
\author[e]{and Ville Vaskonen}

\affiliation[a]{Centro de Investigaciones, Universidad Antonio Nari\~{n}o,\\
Carrera 3 Este \# 47A-15, Bogot\'{a}, Colombia}
\affiliation[b]{Faculty of Physics, University of Warsaw,\\Pasteura 5, 02-093 Warsaw, Poland}
\affiliation[c]{University of Helsinki and Helsinki Institute of Physics, \\
                      P.O.~Box 64, FI-00014, Helsinki, Finland}
\affiliation[d]{Astronomy Unit, Queen Mary University of London, \\ Mile End Road, London, E1 4NS, United Kingdom}
\affiliation[e]{National Institute of Chemical Physics and Biophysics, \\ R\"avala 10, 10143 Tallinn, Estonia}
                      
\emailAdd{nicolas.bernal@uan.edu.co}
\emailAdd{matti.heikinheimo@helsinki.fi}
\emailAdd{t.tenkanen@qmul.ac.uk}
\emailAdd{kimmo.i.tuominen@helsinki.fi}
\emailAdd{ville.vaskonen@kbfi.ee}

\abstract{We present an overview of scenarios where the observed Dark Matter (DM) abundance consists of Feebly Interacting Massive Particles (FIMPs), produced non-thermally by the so-called {\it freeze-in} mechanism. In contrast to the usual freeze-out scenario, frozen-in FIMP DM interacts very weakly with the particles in the visible sector and never attained thermal equilibrium with the baryon-photon fluid in the early Universe. Instead of being determined by its annihilation strength, the DM abundance depends on the decay and annihilation strengths of particles in equilibrium with the baryon-photon fluid, as well as couplings in the DM sector. This makes frozen-in DM very difficult but not impossible to test. In this review, we present the freeze-in mechanism and its variations considered in the literature ({\it dark freeze-out} and {\it reannihilation}), compare them to the standard DM freeze-out scenario, discuss several aspects of model building, and pay particular attention to observational properties and general testability of such feebly interacting DM.
}


\begin{document}
\begin{flushright}
PI/UAN--2017--602FT\linebreak HIP--2017--08/TH
\end{flushright}
\maketitle

%
\section{Introduction}

There is compelling evidence for the existence of Dark Matter (DM), an unknown, non-baryonic matter component whose abundance in the Universe exceeds the amount of ordinary matter roughly by a factor of five~\cite{Ade:2015xua}.
Still, the non-gravitational nature of DM remains a mystery~\cite{Bergstrom:2000pn, Bertone:2016nfn, deSwart:2017heh}. The fact that the observed abundances of dark and visible matter are in the same ballpark may be considered as indicative that both forms of matter have a common origin and they were once in thermal equilibrium with each other.
In the previous decades, this class of scenarios has received by far the biggest attention, both theoretically and experimentally.
Most prominent in this class are extensions of the Standard Model of particle physics (SM) that feature Weakly Interacting Massive Particles (WIMPs) as DM.
WIMPs typically carry electroweak scale mass and couple to the SM with a strength that is reminiscent to that of the weak interactions. The observation that this theoretical setup gives the observed relic abundance is the celebrated \textit{WIMP miracle}~\cite{Arcadi:2017kky}. Another appealing ingredient of the WIMP paradigm is the fact that the same interaction that determines the DM abundance is also responsible for making the paradigm testable by terrestrial experiments.

Despite the fact that DM has been searched for decades, the studies have yielded no overwhelming evidence for what DM actually is. A crucial challenge to the WIMP DM paradigm is the lack of a confirmed experimental detection signal.
The worldwide program for detecting WIMP DM using a multi-channel and multi-messenger
approach has followed three main strategies:
$i)$ direct detection of WIMP-nucleon scattering, $ii)$ indirect detection of WIMPs by measuring the products of their annihilation or decay into SM particles, and
$iii)$ detection of WIMPs produced at high energy colliders such as the LHC. For some recent reviews and reports, see e.g. Refs.~\cite{Baer:2014eja, Klasen:2015uma, Adhikari:2016bei, Alexander:2016aln, Tulin:2017ara}.

From cosmology, we know that to produce the observed amount of light elements and retain the observed peak structure of the Cosmic Microwave Background (CMB) radiation spectrum, the SM particles must have been in state of thermal equilibrium and their energy density must have dominated the evolution of the Universe from the time of big bang nucleosynthesis (BBN) at $T_{\rm BBN}\simeq 4$~MeV to the matter-radiation equality at $T_{\rm eq}\simeq 0.8$~eV~\cite{Kawasaki:2000en, Hannestad:2004px, Ichikawa:2005vw, DeBernardis:2008zz, Ade:2015xua}. On the other hand, DM may or may not have been part of the same heat bath.

In the standard WIMP paradigm, DM is a thermal relic produced by the {\it freeze-out} mechanism. However, the observed DM abundance may have been generated also out of equilibrium by the so-called {\it freeze-in} mechanism~\cite{McDonald:2001vt,Choi:2005vq,Kusenko:2006rh, Petraki:2007gq,Hall:2009bx}. In this scenario, the DM particle couples to the visible SM sector very weakly, so that it never entered chemical equilibrium. Instead, the DM particles were produced by decay or annihilation processes from the visible sector, until the production ceased due to the cooling of the photon temperature below the relevant mass scale connecting the DM particle to the visible sector. Due to the small coupling strength, the DM particles produced via the freeze-in mechanism have been called Feebly Interacting Massive Particles (FIMPs)~\cite{Hall:2009bx}. 

The weakness of interactions between the DM and the SM particles in the freeze-in scenario implies that these models are inherently very difficult to search for in direct detection or collider experiments. However, as the experimental constraints from these experiments are beginning to rule out large parts of the parameter space of the typical WIMP models, the freeze-in scenario has begun to look more appealing. Furthermore, as we shall discuss below, despite the feeble coupling between DM and the SM particles, this scenario is not completely void of experimental signatures to search for.

The FIMP scenario therefore introduces a perfectly adequate mechanism for explaining one of the biggest mysteries in modern physics, the problem of what are the properties of DM and how was its observed abundance generated in the early Universe. As the freeze-in mechanism is gaining popularity, in this paper we provide for the first time a summary of research conducted in the field, introduce in a broader context several interesting phenomena related to DM production that have not been studied exhaustively, and discuss observational properties and testability of very weakly interacting DM in general.

The paper is organized as follows: in Section~\ref{observationalstatus}, we review the current experimental status of DM and discuss several different observational aspects from collider, direct and indirect searches to astrophysical signatures arising from self-interacting DM. Then, in Section~\ref{DMgenesis}, we discuss the DM genesis by first presenting the two basic mechanisms for producing the observed DM abundance, the standard freeze-out and freeze-in scenarios, and then the most important scenarios for how the DM abundance can change after initially produced by decays and annihilations from the visible sector, the {\it dark freeze-out} and {\it reannihilation}. In Section~\ref{models}, we briefly consider examples of studies on the freeze-in mechanism by discussing generic portal models, sterile neutrinos, asymmetric freeze-in, and also other models. In Section~\ref{observations}, we discuss the observational properties of feebly interacting DM, paying particular attention to astrophysical and cosmological signatures. In Section~\ref{conclusions}, we summarize.

\section{Observational Status of Dark Matter}
\label{observationalstatus}

We begin by reviewing the current observational status of DM. There is currently a worldwide program for detecting DM but only a little has been revealed about the concrete micro-physical properties of particle DM, as so far only the gravitational interactions of DM have been confirmed. In this Section we discuss the results from direct and indirect searches and collider experiments, as well as astrophysical signatures arising from DM self-interactions.

\subsection{Direct Detection}

The aim of direct detection experiments is to measure the small recoil energy of a nucleus in an underground detector after a collision with a DM particle gravitationally bound to our galactic halo.
The current status of direct detection searches is ambiguous with a few experiments reporting hints for a DM signal like DAMA/LIBRA~\cite{Bernabei:2010mq, Bernabei:2013xsa}, CoGent~\cite{Aalseth:2010vx, Aalseth:2011wp, Aalseth:2012if, Aalseth:2014eft, Aalseth:2014jpa} and CDMS-II~\cite{Agnese:2013rvf}.
However, such observations seem at face value inconsistent among each other and are typically viewed to be in conflict with the null results of many other experiments~\cite{Aprile:2011hi, Angle:2011th, Felizardo:2011uw, Archambault:2012pm, Ahmed:2012vq, Behnke:2012ys, Aprile:2012nq, Agnese:2014aze, Amole:2015lsj, Agnese:2015ywx, Angloher:2015ewa, Agnese:2015nto, Akerib:2015rjg, Tan:2016zwf, Akerib:2016vxi, Aprile:2016swn, Aprile:2017iyp}.
Currently the XENON1T experiment~\cite{Aprile:2017iyp} places the strongest exclusion limit in the plane of spin-independent DM-nucleon cross section and WIMP mass for large DM masses, while LUX~\cite{Akerib:2016vxi} and PandaX-II~\cite{Tan:2016zwf} have recently reported competitive null results.
The most stringent direct detection constraint to date on the WIMP-proton spin-dependent cross section comes from the PICO-60~\cite{Amole:2017dex} and the LUX~\cite{Akerib:2017kat} experiments (and below $\sim 4$~GeV from the PICASSO experiment~\cite{Behnke:2016lsk}).
A new generation of direct detection DM experiments includes LZ~\cite{Akerib:2015cja, Szydagis:2016few}, DARWIN~\cite{Schumann:2015cpa}, DarkSide-20k~\cite{Agnes:2015ftt, Davini:2016vpd}, PICO-250~\cite{Pullia:2014vra} and SuperCDMS~\cite{Agnese:2016cpb}.

In light of null results from WIMP direct detection experiments, there is considerable interest in exploring new regions of DM parameter space with a variety of different techniques~\cite{Alexander:2016aln}.
Multiple directions have recently been suggested for the detection of elastic and inelastic scatterings of DM in the mass range from keV to MeV:
recently proposed experiments with sensitivity to DM-electron scattering include Refs.~\cite{Essig:2011nj, Essig:2012yx, Lee:2015qva, Essig:2015cda, Green:2017ybv, Emken:2017erx, Essig:2017kqs, Kadribasic:2017obi}. Additional possible methods for detecting sub-GeV DM include semiconductors~\cite{Graham:2012su, Hochberg:2016sqx}, superconductors~\cite{Hochberg:2015pha, Hochberg:2015fth, Hochberg:2016ajh}, scintillating targets~\cite{Derenzo:2016fse}, superfluid helium~\cite{Schutz:2016tid, Knapen:2016cue}, chemical-bond breaking~\cite{Essig:2016crl}, two-dimensional targets~\cite{Capparelli:2014lua, Hochberg:2016ntt, Cavoto:2017otc}, color center production in crystals~\cite{Budnik:2017sbu}, molecular spectroscopy~\cite{Fichet:2017bng}, as well as search for bremsstrahlung radiation from a recoiling nucleus~\cite{Kouvaris:2016afs, McCabe:2017rln}.

\subsection{Indirect Detection}

Indirect detection aims at detecting the flux of final stable particles, mainly electrons, positrons, protons, antiprotons and photons, produced by DM annihilation or decay.
Among those, photons are considered the golden channel for the identification of a possible DM signal, since they are not accelerated by magnetic fields in the galactic environment, and thus preserve the spectral and spatial information of the source signal itself.

Several potential signals have appeared in indirect DM searches over the past few years.
Most notably, in the case of the Galactic center gamma ray excess in the GeV energy range~\cite{Goodenough:2009gk, Abazajian:2012pn, Daylan:2014rsa, Zhou:2014lva, Calore:2014xka, TheFermi-LAT:2015kwa, Linden:2016rcf}, an anomalous emission of gamma rays coming from the inner region of the Galaxy has been reported.
Various interpretations of this excess have been put forward,  from astrophysical processes (e.g. Refs.~\cite{Petrovic:2014uda, Calore:2014oga, Cholis:2014lta, Petrovic:2014xra, Cholis:2015dea,Macias:2016nev}) to DM annihilation (e.g. Refs.~\cite{Calore:2014nla, Bertone:2015tza}).
Although there is strong support for the GeV excess to arise from a population of faint, unresolved point sources, DM interpretation has not been yet robustly excluded.
However, this interpretation is challenged by the latest constraints from dwarf spheroidal galaxies~\cite{Fermi-LAT:2016uux}.

An anomalous emission line at the 3.55~keV energy from X-ray images taken by the telescope XMM-Newton~\cite{Struder:2001bh, Turner:2000jy} has been found in a stacked analysis of the Perseus cluster,
Coma-Centaurus-Ophiuchus clusters and 69 other clusters~\cite{Bulbul:2014sua}.
Further evidence has been gathered also from other galaxy clusters~\cite{Ruchayskiy:2015onc, Iakubovskyi:2015dna, Bulbul:2014ala, Franse:2016dln}.
However, the status of this line is controversial.
In fact, these results are in tension with the Suzaku data~\cite{Urban:2014yda}, which show no indication of a 3.55~keV line in the Coma, Virgo and Ophiuchus clusters.
On top of that, these findings are in tension with the Hitomi observations of the Perseus cluster, which do not support an excess in this cluster~\cite{Aharonian:2016gzq}.
Nevertheless, the line has been recently confirmed in the summed data from deep Chandra blank fields, the Chandra Deep Field South and COSMOS~\cite{Cappelluti:2017ywp}.
This spectral feature can have a DM related origin, but it also may have no connection with the DM sector and come from some atomic transition (of a potassium origin~\cite{Jeltema:2014qfa}) or from some systematic errors.
However, while the DM interpretation has been challenged by many authors~\cite{Jeltema:2014qfa, Anderson:2014tza, Boyarsky:2014paa, Urban:2014yda, Gu:2015gqm, Jeltema:2015mee, Ruchayskiy:2015onc, Bulbul:2016yop, Aharonian:2016gzq}, it is not in contradiction with observations of dwarf galaxies~\cite{Malyshev:2014xqa} or the galactic center~\cite{Riemer-Sorensen:2014yda}.

The AMS-02 experiment on board the international space station has measured the fluxes of various charged particles in cosmic rays. The ratio of antiproton and proton fluxes exhibits an excess over the estimated background~\cite{Aguilar:2016kjl}, that has been interpreted as a signal of DM annihilations~\cite{Cuoco:2016eej, Cui:2016ppb, Huang:2016tfo, Li:2016uph, Feng:2017tnz, Jin:2017iwg, Jia:2017kjw, Cuoco:2017rxb}. Intriguingly, some of the above references report good agreement with DM annihilations as an explanation for both the AMS antiproton data and the galactic center gamma ray excess discussed above.
However, within systematic errors, secondary astrophysical production alone can account for the data~\cite{Giesen:2015ufa}.
Furthermore, a high-energy cosmic positron excess has been discovered by PAMELA~\cite{Adriani:2008zr} and confirmed by AMS-02~\cite{Aguilar:2013qda}.
While it can be produced by astrophysical objects like pulsars or supernova remnants (see e.g. Refs.~\cite{Hooper:2008kg, DiMauro:2014iia}), DM annihilations can also be involved (see e.g. Refs.~\cite{Bai:2009ka, Chen:2015cqa}).

\subsection{Collider Searches}
\label{sec:collsearches}

As with the direct detection, DM has to interact sufficiently strongly with the SM particles to yield observable signatures at colliders. The efforts to search for DM in collider experiments are mainly focused on traditional WIMPs, such as the neutralino.
These particles are searched for in model-dependent scenarios that constrain the parameter space of UV-complete models, effective field theories and simplified models~\cite{Goodman:2011jq, Abdallah:2014hon, Malik:2014ggr, Abdallah:2015ter, Abercrombie:2015wmb, Choudhury:2015lha, Kahlhoefer:2015bea, Englert:2016joy}, as well as in more model-independent searches for missing energy accompanied by a monophoton~\cite{Khachatryan:2014rwa, Aad:2014tda, Khachatryan:2016hns, Aaboud:2016uro, ATLASCollaboration:2016wlb, Khachatryan:2016ojf}, a mono-$Z$~\cite{Aad:2013oja, Aad:2014vka}, a monojet~\cite{Shoemaker:2011vi, An:2012va, Aad:2014nra, Khachatryan:2014rra, Aad:2015zva, Khachatryan:2015wza, Barducci:2016fue} or a mono-Higgs~\cite{Berlin:2014cfa, Aad:2015yga, No:2015xqa, Aad:2015dva, Aaboud:2016obm, Sirunyan:2017hnk}. Constraints from the monophoton searches at BaBar~\cite{Aubert:2008as, Essig:2013vha} are stringent at lower masses, between a few hundred MeV and $\sim 10$~GeV.

Recently, the use of simplified models, that typically contain a mediator particle connecting the DM sector to the SM, has been under a lot of focus in DM collider searches, as opposed to the use of effective field theories, containing non-renormalizable effective operators that are obtained in the limit of integrating out the heavy mediator particle. The contact interaction limit of effective non-renormalizable operators is only valid when the mediator mass is safely above the center of mass energy of the hard scattering event, and this assumption does not generally hold in the 13 TeV LHC. This observation has three consequences: First, the presence of the mediator propagator has to be taken into account in predictions for the DM production signal, as it can affect the kinematical characteristics of the events compared to the contact interaction limit. Second, the production of the mediator particle may also result in final states other than a DM particle pair, and these visible decay modes can be used to constrain such models.
Third, it can alter the interplay between cosmological and collider constraints.
See Refs.~\cite{Albert:2017onk, Banerjee:2017wxi} and references therein for related discussion.

\subsection{Dark Matter Self-interactions}
\label{DMSI}

A new strategy for DM searches consists of looking for effects associated with possible non-gravitational DM self-interactions in the sky.
The cosmological concordance model ($\Lambda$CDM) presents DM as a cold, collisionless fluid. However, evidence has been accumulating since the 1990's, suggesting that gravitational N-body simulations performed with cold DM only, while extremely successful at large scales, do not correctly reproduce the observed structure at the galactic and sub-galactic scales. Two long-standing puzzles of the collisionless cold DM paradigm are the `cusp vs. core'~\cite{Moore:1994yx, Flores:1994gz, Navarro:1996gj, deBlok:2009sp, Oh:2010mc, Walker:2011zu} and the `too-big-to-fail'~\cite{BoylanKolchin:2011de, Garrison-Kimmel:2014vqa} problems.
These issues are collectively referred to as small scale structure problems of the $\Lambda$CDM model; for a recent review, see Ref.~\cite{Tulin:2017ara}.

While it has been proposed that the above tensions could be addressed by carefully accounting for the effects of baryonic physics~\cite{MacLow:1998djk, Governato:2009bg, Silk:2010aw, VeraCiro:2012na, Sawala:2015cdf, Fattahi:2016nld}, alternative solutions are offered by warm~\cite{SommerLarsen:1999jx, Bode:2000gq}, decaying~\cite{Wang:2014ina} or self-interacting DM~\cite{Tulin:2017ara}. Thus, the small scale structure of the Universe can be seen as a probe for the properties of DM, which can potentially distinguish between the standard picture of cold collisionless and collisional DM. It is noteworthy that this is independent of how strongly DM interacts with the SM particles and can, in principle, allow to efficiently probe DM even if the other probes such as collider or direct detection experiments fail to identify properties of DM.

These tensions can be alleviated if at the scale of dwarf galaxies DM exhibits a large self-scattering cross section, $\sigma$, over DM particle mass, $m$, in the range $0.1\lesssim\sigma/m\lesssim10$~cm$^2/$g~\cite{Spergel:1999mh, Wandelt:2000ad, Buckley:2009in, Vogelsberger:2012ku, Rocha:2012jg, Peter:2012jh, Zavala:2012us, Vogelsberger:2014pda, Elbert:2014bma, Kaplinghat:2015aga}.
Nevertheless, the non-observation of an offset between the mass distribution of DM and galaxies in the Bullet Cluster constrains such self-interacting cross section, concretely $\sigma/m<1.25$ cm$^2$/g at $68\%$~CL~\cite{Clowe:2003tk, Markevitch:2003at, Randall:2007ph}, i.e. around $10^{12}$~pb for a DM of 1~GeV mass.
Similarly, recent observations of cluster collisions lead to the constraint $\sigma/m<0.47$~cm$^2$/g at $95\%$~CL~\cite{Harvey:2015hha}. However, some of the methods and assumptions utilized to obtain these results have been questioned in the recent literature~\cite{Robertson:2016xjh, Kim:2016ujt, Robertson:2016qef, Wittman:2017gxn} and the actual limit could be less stringent.

Finally, observations of the galaxy cluster Abell 3827 suggest an offset between the stars and the DM halo in at least one of the four central galaxies.
If interpreted solely as an effect of DM self-interaction, it was argued to imply a non-vanishing $\sigma/m$ of the order of $10^{-4}$~cm$^2$/g~\cite{Massey:2015dkw}. However, these results have been reinterpreted using an improved kinematical analysis, obtaining $\sigma/m\sim 1.5$~cm$^2$/g in the case of contact interactions~\cite{Kahlhoefer:2015vua, Taylor:2017ipx}.
The DM substructure observed in Abell 520~\cite{Mahdavi:2007yp, Jee:2014hja}, coincident with the hot gas and not hosting any stars, has been interpreted as suggestive of self-interacting DM in Refs.~\cite{Heikinheimo:2015kra,Sepp:2016tfs}.

The self-scattering of DM particles leads to heat transfer that
decreases the density contrast in the centers of DM halos turning
cusps into cores and changing the subhalo abundance matching due to a
lower halo concentration. Self-interacting DM therefore directly
addresses the two small-scale
problems.  Although this effect alone cannot efficiently
reduce the formation rate of luminous galaxies in DM subhalos, it may
still alleviate the `missing satellites'
problem~\cite{Klypin:1999uc, Moore:1999nt} with the help of additional DM physics (e.g. warm or decaying DM) or baryonic feedback.
While latest observations tend to prefer the latter option~\cite{Bechtol:2015cbp, Drlica-Wagner:2015ufc}, the problem remains unsettled~\cite{Koposov:2007ni, Jethwa:2016gra,Tulin:2017ara}.

\vspace{0.5cm}

In this Section we have reviewed the global program for detecting a WIMP DM particle. Despite the huge experimental and observational effort, there is no compelling non-gravitational evidence for a DM particle.
To progress, it is necessary to make use of the complementarity between different experiments and different detection techniques~\cite{Mena:2007ty, Drees:2008bv, Bernal:2008zk, Bernal:2008cu, Bergstrom:2010gh, Pato:2010zk, Arisaka:2011eu, Cerdeno:2013gqa, Arina:2013jya, Peter:2013aha, Kavanagh:2014rya, Roszkowski:2016bhs, Roszkowski:2017dou} in order to ameliorate
determination of the particle physics parameters and disentangle possible degeneracies.
Furthermore, one has to take into account astrophysical uncertainties~\cite{Green:2002ht, Zemp:2008gw, McCabe:2010zh, Pato:2010yq, Fairbairn:2012zs, Bernal:2014mmt, Bernal:2015oyn, Bernal:2016guq, Benito:2016kyp, Green:2017odb} when interpreting the results of the DM searches.
It is certainly still possible that an experimental signature of WIMP DM will arise in the coming years, but as the natural parameter space of these models is closing~\cite{Arcadi:2017kky}, we must keep an open mind and extend our search into new realms beyond the standard WIMP paradigm.


\section{Dark Matter Genesis}
\label{DMgenesis}

In this Section discuss how the observed DM abundance can arise in the early Universe. We first present the two basic mechanisms for producing the observed DM abundance, the standard {\it freeze-out} and its recently proposed alternative, the {\it freeze-in} scenario, and then the most important scenarios for how the DM abundance can change after initially produced by freeze-in type processes, the {\it dark freeze-out} and {\it reannihilation}. A brief summary of all these mechanisms is given in Section \ref{DMgensummary}.

Depending on the coupling between DM and particles in the visible sector, DM either reached thermal equilibrium with the visible sector in the early Universe, or did not. In case it did, the DM genesis is based on the standard {\it freeze-out} mechanism. In the freeze-out mechanism the comoving DM particle number density freezes to a constant value roughly when the interaction rate between the visible sector and DM particles becomes smaller than the Hubble rate. However, if the coupling is small enough, DM was never in thermal equilibrium with the visible sector. In that case, the DM abundance can, in the simplest case, be produced by the {\it freeze-in} mechanism\footnote{We define freeze-in to refer to a mechanism where the DM density is produced non-thermally by decays and annihilations of particles in thermal equilibrium. This excludes several other mechanisms for generating the DM abundance non-thermally such as incomplete reheating~\cite{Kofman:1994rk}, asymmetric reheating~\cite{Hodges:1993yb, Berezhiani:1995am, Adshead:2016xxj} and Dodelson-Widrow mechanism~\cite{Dodelson:1993je}.} where the comoving DM number density freezes to a constant value when the number densities of particles producing DM become negligible due to Boltzmann suppression. In the simplest case the freeze-out occurs by $2\to2$ DM annihilations, whereas freeze-in production is dominated by $1\to2$ decays or $2\to2$ annihilations of visible sector particles into DM. In more complicated model setups interactions in the DM sector can further modify the DM number density once it has frozen in/out from the visible sector. We will discuss these scenarios in Sections~\ref{freezeout}, \ref{freezein} and~\ref{dark_FO}.

To accurately solve for the DM relic density, one must follow the microscopic evolution of particle phase space distribution functions. In the following we will assume, for simplicity, that there is only one DM particle and denote it by $\chi$. Time evolution of the phase space distribution function $f_\chi(p,\,t)$ of the DM particle $\chi$ is then given by the Boltzmann equation\footnote{Notice that this is an approximation which neglects thermal effects. These may reduce the abundance of non-thermally produced DM by $\mathcal{O}(10-100)\%$~\cite{Hamaguchi:2011jy}.}~\cite{Kolb:1990vq}
\be \label{DM_boltzmann}
\left( \partial_t - H\,\vec{p}_\chi\,\cdot \nabla_{p}  \right) f_\chi(p_\chi,\,t) = \sum_j \frac{1}{E_\chi}\int \mathrm{d}C_{\chi,\,j}\,,
\ee
where the sum is over all processes involved, $\vec{p}_\chi$ is the physical momentum and $E_\chi$ is the energy of $\chi$ particles, $t$ is the cosmic time, and $H$ corresponds to the Hubble parameter related to the total energy density via the Friedmann equation, $\rho = 3H^2 M_{\rm P}^2/(8\pi)$,
where $M_{\rm P}$ is the Planck mass. Due to isotropy of the Universe, $f(p_\chi,t)$ depends only on the magnitude of $\vec{p}_\chi$. The differential collision term $\mathrm{d}C_\chi$ for the process $\chi + a_1+ a_2 + \dots + a_n \to b_1 + b_2 + \dots + b_m$ (or, equivalently, to $b_1 + b_2 + \dots + b_m \to \chi + a_1+ a_2 + \dots + a_n$ with a different sign) is
\bea
\mathrm{d}C_\chi \equiv& - \mathrm{d}\Pi_{a_1} \mathrm{d}\Pi_{a_2} \dots \mathrm{d}\Pi_{b_1} \mathrm{d}\Pi_{b_2} \dots \\
&\quad\times (2\pi)^4\,\delta^4\left(p_\chi + p_{a_1} + \dots - p_{b_1} - p_{b_2} - \dots \right) \\
&\quad\times \left[ |M|_{\chi + a_1 + \dots \to b_1 + b_2 + \dots}^2 f_\chi f_{a_1} \dots (1\pm f_{b_1}) (1\pm f_{b_2}) \dots \right. \\
&\qquad- \left. |M|_{b_1 + b_2 + \dots \to \chi + a_1 + \dots}^2 f_{b_1} f_{b_2} \dots (1\pm f_\chi) (1\pm f_{a_1}) \dots \right],
\eea
where $+$ applies for bosons and $-$ for fermions, $|M|_{\chi + a_1 + \ldots \to b_1 + b_2 + \ldots}^2$ and $|M|_{b_1 + b_2 + \ldots \to \chi + a_1 + \ldots}^2$ are the squared transition amplitudes, averaged over both the initial and final states, and 
\be
\mathrm{d}\Pi_k \equiv \frac{g_k}{(2\pi)^3} \frac{\mathrm{d}^3p_k}{2 E_k}\,,
\ee
is the usual phase space measure. Here the number of intrinsic degrees of freedom of particle $k$ is denoted by $g_k$, its momentum and energy by $p_k$ and $E_k$, respectively, and its phase space distribution by $f_k$. 

In general, Boltzmann equations form a coupled set of integro-differential equations for distribution functions of all particle species involved in DM production, but if all but one species are assumed to remain in thermal equilibrium with each other, solving the DM abundance reduces to a single integro-differential equation for the DM distribution $f_\chi$. Once all interactions and initial conditions are known, the solution of Eq.~\eqref{DM_boltzmann} gives the present-day DM abundance.

\subsection{Summary of Mechanisms}
\label{DMgensummary}

\begin{figure}
\begin{center}
\includegraphics[width = 0.6\textwidth]{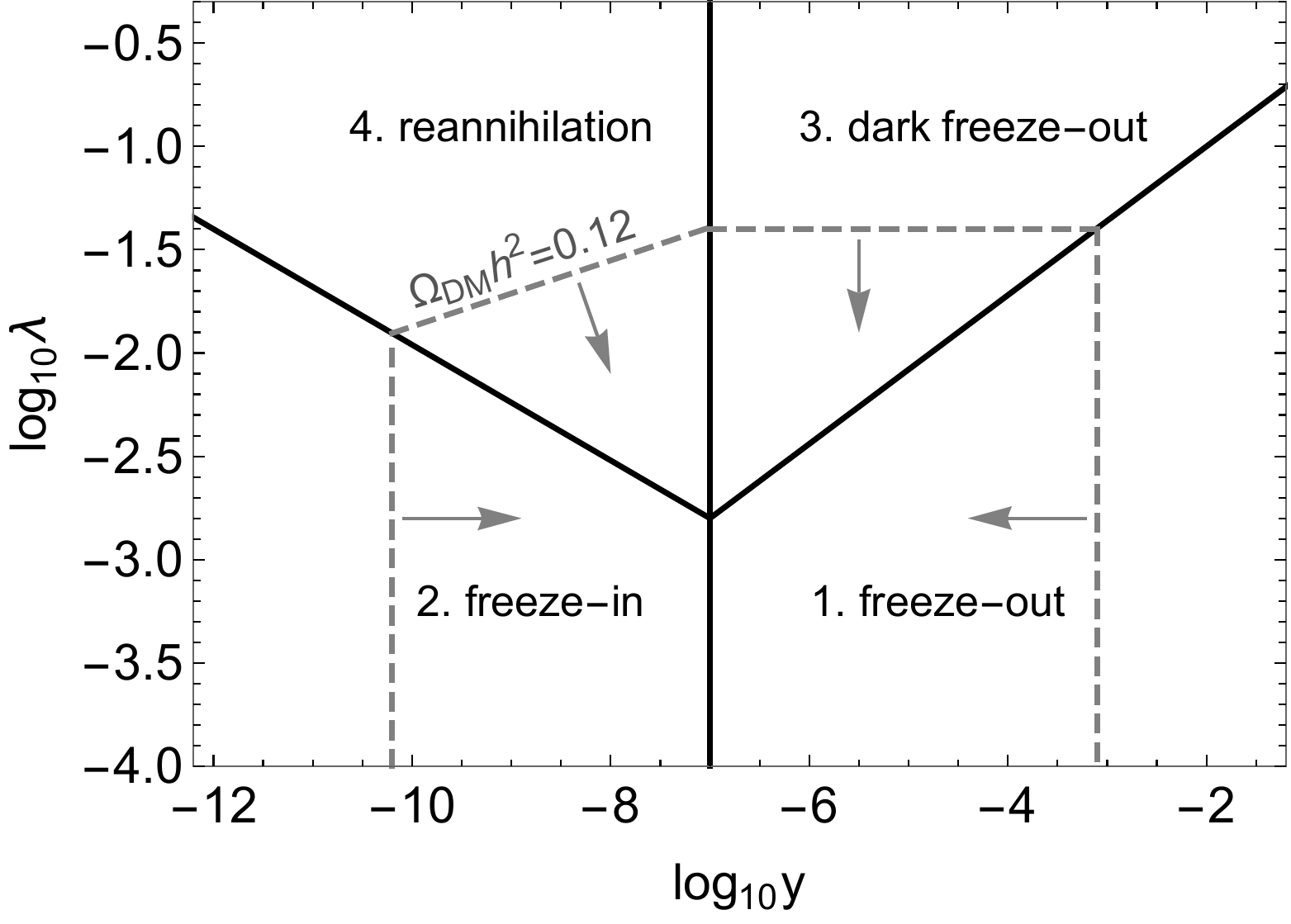}
\caption{A schematic representation of the regions where each of the four mechanisms described in the text dominate. The observed DM relic abundance is obtained on the dashed gray line, and the arrows show the gradient for DM abundance. The parameters $y$ and $\lambda$ are the DM coupling to the visible sector and the DM self-coupling, respectively.}
\label{fig:phases}
\end{center}
\end{figure}

As will be described in detail in the following Sections, there are in total four distinct mechanisms for producing the observed DM abundance through ordinary decay and scattering processes, three of them including an initial freeze-in type process.\footnote{A possible caveat to this statement is the fact that the DM sector may have been populated already by reheating in the early Universe~\cite{Chu:2011be, Bernal:2015ova,Tenkanen:2016jic}. If that was the case, the DM abundance may undergo dark freeze-out without ever been sourced by or having been in contact with the visible sector heat bath.} We summarize them here:
\begin{enumerate}
\item {\bf Freeze-out mechanism}: DM decouples from the visible sector heat bath when its interactions with the bath particles cannot compete with the expansion of the Universe anymore and its comoving number density freezes to a constant value. This requires roughly a weak scale coupling, $y\simeq\mathcal{O}(0.1)$, although the value is highly model-dependent.
\item {\bf Freeze-in mechanism}: DM was never in thermal equilibrium with the visible sector, and the comoving DM abundance freezes to a constant value when the number densities of visible sector particles producing DM either by decays or annihilations become Boltzmann-suppressed, ending the yield. This requires usually a very small coupling, $y\simeq\mathcal{O}(10^{-7})$ or less, to prevent the dark sector from thermalizing with the visible sector and to obtain the correct relic abundance.
\item {\bf Dark freeze-out mechanism}: DM never became in thermal equilibrium with the visible sector but comprised an equilibrium heat bath within its own dark sector, which was initially populated by a freeze-in-type yield from the visible sector. In this scenario DM does not annihilate into visible sector particles but into states of the dark sector. The dynamics are similar to the usual freeze-out scenario except for the fact that the two sectors may have different temperatures.
\item {\bf Reannihilation mechanism}: A scenario where the dark sector thermalizes within itself but where the dark freeze-out would occur already before the yield from the visible sector has ended. The on-going particle production from the visible sector keeps increasing the DM relic abundance and thus resumes annihilations of dark sector particles, forcing the comoving number density of DM to freeze out only after the yield has ended.
\end{enumerate}
In Fig.~\ref{fig:phases}, the dashed gray line illustrates the values of the DM coupling to the visible sector $y$ and the DM self-coupling $\lambda$ for which different DM production mechanisms are realized in a scenario where the dark sector consists of the DM particle only (see also Fig.~3 of Ref.~\cite{Chu:2011be} and Fig.~4 of Ref.~\cite{Bernal:2015ova}). Next, we will discuss all these mechanisms one by one.

\subsection{Freeze-out}
\label{freezeout}

To better understand properties of the freeze-in mechanism, we first review the DM production mechanism most studied in the literature: the freeze-out scenario. In that case DM rapidly reaches thermal equilibrium with the visible sector, and the relic abundance is determined when the interactions between the DM and visible sector particles cannot compete with the expansion of the Universe anymore.

In the simplest case where DM undergoing thermal freeze-out consists of stable particles $\chi$ with no initial asymmetry between particles and anti-particles nor large self-interactions, the abundance can be solved simply by considering annihilation and inverse annihilation processes, $\chi\bar{\chi} \leftrightarrow {\rm SM}$. For asymmetric DM and the effect of large self-interactions, see e.g. Ref.~\cite{Petraki:2013wwa, Zurek:2013wia} and Refs.~\cite{Carlson:1992fn, Hochberg:2014dra, Pappadopulo:2016pkp, Farina:2016llk}, respectively.

Integrating Eq.~\eqref{DM_boltzmann} over momentum, one finds that the evolution of the DM number density is governed by~\cite{Zeldovic:1965rys, Lee:1977ua}
\begin{equation}
\label{fo_boltzmann}
\frac{dn_\chi}{dt} + 3\,H\,n_\chi = -\langle v\sigma_\chi \rangle\left[n_\chi^2 - (n_\chi^{\rm eq})^2\right]\,,
\end{equation}
where $\langle v\sigma_\chi\rangle$ is the thermally-averaged total DM annihilation cross section times velocity~\cite{Gondolo:1990dk}, $n_\chi$ is the actual DM number density and $n_\chi^{\rm eq}$ is the DM equilibrium number density.

By defining the commonly used dimensionless variables $Y\equiv n_\chi/s$ and $x\equiv m_\chi/T$, where $s$ and $T$ are the entropy density and temperature of the visible sector heat bath, respectively, and $m_\chi$ is the DM particle mass, and assuming the number of relativistic degrees of freedom remains constant, Eq.~\eqref{fo_boltzmann} can be cast in the form
\begin{equation}
\frac{x}{Y_{\rm eq}}\frac{dY}{dx} = -\frac{\Gamma}{H}\left[\left(\frac{Y}{Y_{\rm eq}}\right)^2-1\right]\,,
\end{equation}
where $\Gamma \equiv n_\chi^{\rm eq}\langle v\sigma_\chi\rangle$ is the effective interaction rate between DM and bath particles and $Y_{\rm eq}$ is proportional to the number of DM particles in equilibrium.
Exceptions to the vanilla DM freeze-out are the coannihilation, the annihilation into forbidden channels and the annihilation near poles~\cite{Griest:1990kh}; as well as the semi-annihilations~\cite{Hambye:2008bq, DEramo:2010keq, Belanger:2012vp} and the co-scatterings~\cite{DAgnolo:2017dbv, Garny:2017rxs}.

\begin{figure}
\begin{center}
\includegraphics[width=.48\textwidth]{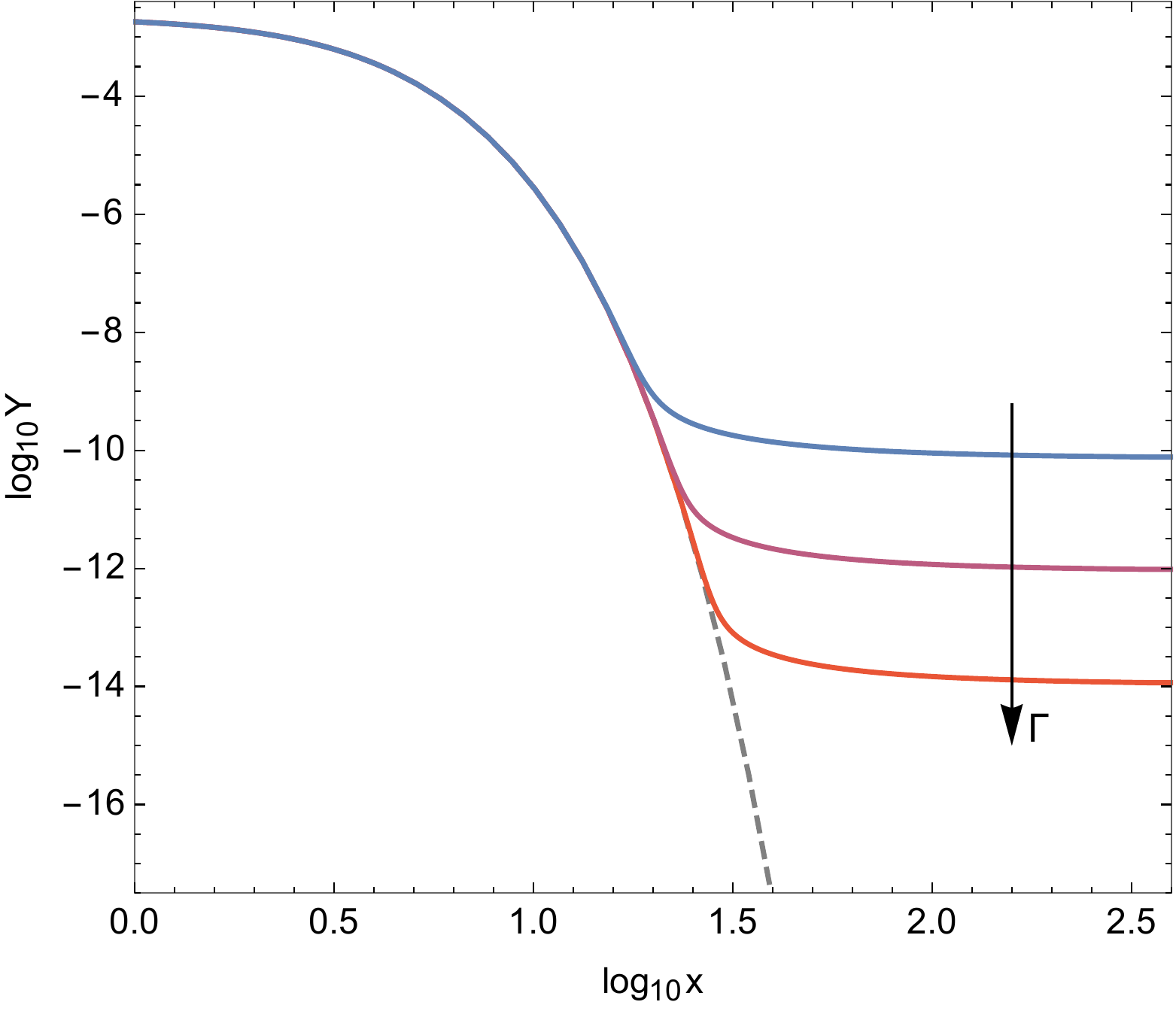} \hspace{2mm}
\includegraphics[width=.48\textwidth]{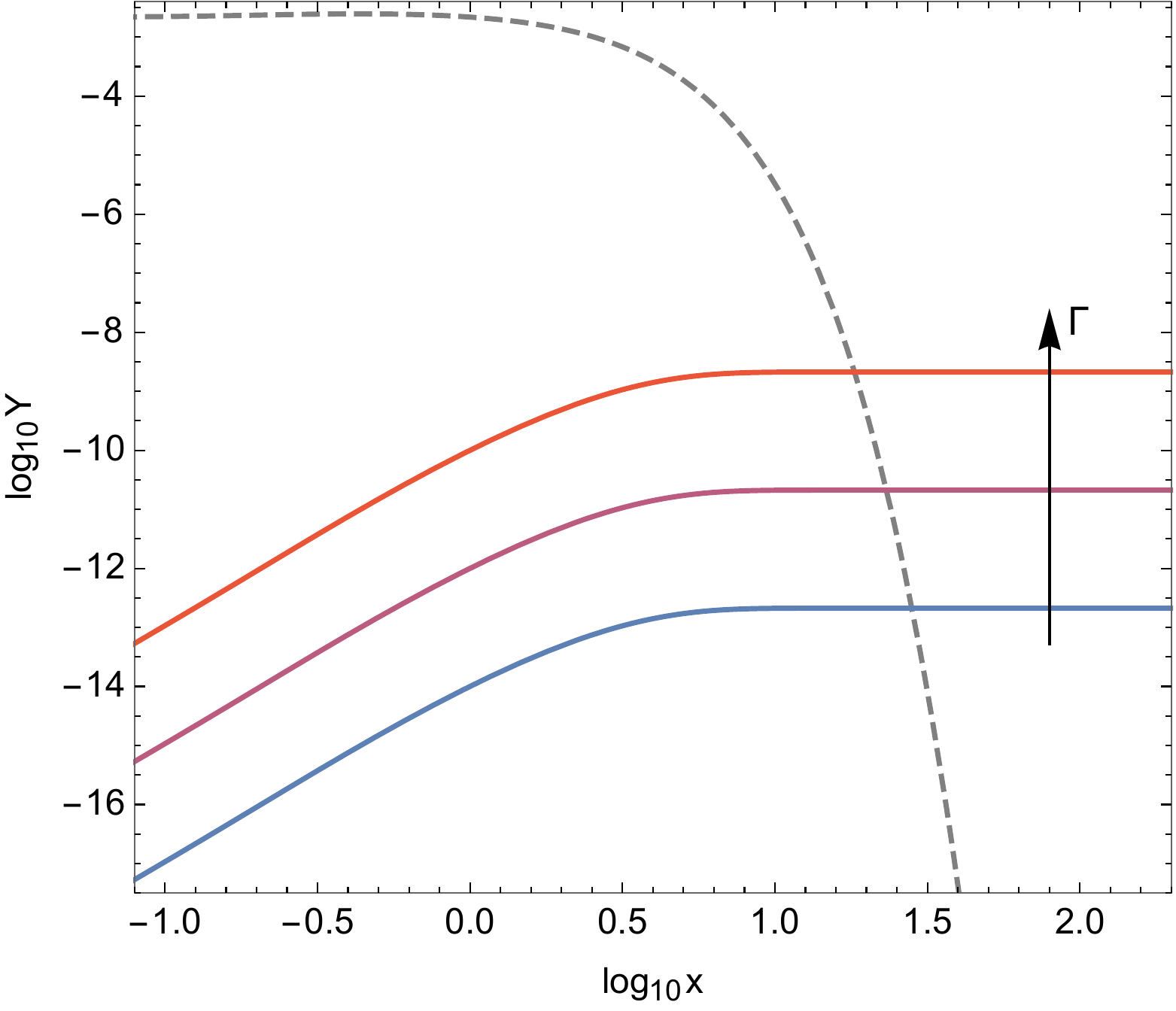}
\caption{The two basic mechanisms for DM production: the freeze-out (left panel) and freeze-in (right panel), for three different values of the interaction rate between the visible sector and DM particles $\chi$ in each case. The arrows indicate the effect of increasing the rate $\Gamma$ of the two processes.  In the left panel $x=m_\chi/T$ and gray dashed line shows the equilibrium density of DM particles. In the right panel  $x=m_\sigma/T$, where $\sigma$ denotes the particle decaying into DM, and the gray dashed line shows the equilibrium density of $\sigma$. In both panels $Y=n_\chi/s$, where $s$ is the entropy density of the baryon-photon fluid.}
\label{FIvsFO}
\end{center}
\end{figure}

The DM freeze-out is depicted in the left panel of Fig.~\ref{FIvsFO}, which shows the evolution of the comoving DM number density for different values of the interaction rate $\Gamma$. As long as $H<\Gamma$, equilibrium between the visible sector and DM particles is maintained. The DM abundance follows the equilibrium number density, $Y\simeq Y_{\rm eq}$, until the annihilations shut off at $H\simeq\Gamma$, and the DM abundance freezes out. After that the comoving DM number density remains constant, $n_\chi\propto a^{-3}$. This is true for scenarios where the so-called DM {\it cannibalization} phase is absent. This refers to a case where number-changing interactions, such as $\chi\chi\chi\to \chi\chi$ or $\chi\chi\chi\chi\to \chi\chi$ scatterings, modify the DM abundance even after it has fallen out of thermal equilibrium with the baryon-photon fluid~\cite{Carlson:1992fn}. In this Section we neglect these processes for simplicity but turn back to this issue in Section~\ref{dark_FO}.

The standard approximate solution for the present-day DM abundance then becomes~\cite{Kolb:1990vq}
\begin{equation}
\label{fo_abundance}
\Omega_\chi h^2 \simeq 5.36\times 10^9 \frac{\sqrt{g_*}}{g_{*s}} \frac{m_\chi}{T_f} \frac{(n+1)\,{\rm GeV}^{-1}}{M_{\rm P}\,\langle v\sigma_\chi\rangle} ,
\end{equation}
where $g_{*s}$ and $g_*$ are the effective numbers of relativistic degrees of freedom related to the entropy and energy densities, respectively, $T_f$ corresponds to the freeze-out temperature, $n=0$ for $s$-wave annihilation, $n=1$ for $p$-wave annihilation, and so on. Here we assumed that the freeze-out occurs when DM is non-relativistic.

Eq.~\eqref{fo_abundance} has an important feature: the present abundance is inversely proportional to the DM annihilation cross section. This can be understood by recalling that in the freeze-out scenario DM particles are initially in thermal equilibrium with the visible sector and the stronger the interaction between them is, the longer the DM particles remain in equilibrium and thus the more their abundance gets diluted before the eventual freeze-out. This can also be seen in the left panel of Fig.~\ref{FIvsFO}.

\subsection{Freeze-in}
\label{freezein}

The above discussion was based on the assumption that the DM initially reached thermal equilibrium with the visible sector. However, if the coupling between the visible sector and DM particles is very small, typically $y\simeq\mathcal{O}(10^{-7})$ or less~\cite{Enqvist:1992va, Enqvist:2014zqa}, interactions between them are not strong enough for DM to reach thermal equilibrium and freeze-out cannot happen. Instead, the observed DM abundance can be produced by the freeze-in mechanism~\cite{McDonald:2001vt, Hall:2009bx}. In this case, the particle undergoing the freeze-in is referred to as a FIMP (Feebly Interacting Massive Particle)~\cite{Hall:2009bx}, as opposed to the WIMP.

In the simplest case, the initial number density of DM particles is either zero or negligibly small, and the observed abundance is produced by bath particle decays, for instance by $\sigma\to \chi\chi$, where $\sigma$ is a particle in the visible sector heat bath~\cite{McDonald:2001vt, Kusenko:2006rh, Petraki:2007gq,Hall:2009bx, Cheung:2010gj, Cheung:2010gk, Yaguna:2011qn, Chu:2011be, Blennow:2013jba, Arcadi:2016dbl, Kaneta:2016wvf}. The freeze-in yield is active until the number density of $\sigma$ becomes Boltzmann-suppressed, $n_{\sigma}\propto {\rm exp}(-m_{\rm \sigma}/T)$. The comoving number density of DM particles $\chi$ then becomes a constant and the DM abundance freezes in. This is depicted in the right panel of Fig. \ref{FIvsFO}.

If the initial DM number density is small compared to the equilibrium number density the backreaction term in the Boltzmann equation~\eqref{DM_boltzmann} can be neglected, and the evolution of $n_\chi$ is given by
\be
\label{FI_boltzmann}
\frac{{\rm d}n_\chi}{{\rm d}t} + 3\,H\,n_\chi = 2\,\Gamma_{\sigma\to \chi\chi} \frac{K_1(m_\sigma/T)}{K_2(m_\sigma/T)} n_\sigma^{\rm eq}\,,
\ee
where $K_j$ are modified Bessel functions of the second kind, $\Gamma_{\sigma\to\chi\chi}$ is the decay width and $n_\sigma^{\rm eq}$ is the equilibrium number density of $\sigma$. In deriving Eq.~\eqref{FI_boltzmann}, we assumed that the $\sigma$ particles obey Maxwell-Boltzmann statistics, $f_\sigma = {\rm exp}(-E_\sigma/T)$. Defining then $Y\equiv n_\chi/s$ and $x\equiv m_\sigma/T$, and assuming again that the number of relativistic degrees of freedom remains constant during DM production, Eq.~\eqref{FI_boltzmann} can be rewritten as
\be
\label{FI_boltzmannY}
\frac{x}{Y_\sigma^{\rm eq}} \frac{\mathrm{d}Y}{\mathrm{d}x} = 2\, \frac{\Gamma_{\sigma\to \chi\chi}}{H} \frac{K_1(x)}{K_2(x)} .
\ee
The approximate solution to this is~\cite{Hall:2009bx}
\begin{equation}
\label{freezeinabundance}
\Omega_{\rm \chi}h^2 \simeq 4.48\times10^{8}\frac{g_\sigma}{g_{*s}\sqrt{g_*}} \frac{m_{\rm \chi}}{\rm GeV} \frac{M_{\rm P}\,\Gamma_{\sigma\to \chi\chi}}{m_{\rm \sigma}^2}\,,
\end{equation}
where $g_\sigma$ is the intrinsic number of degrees of freedom of the $\sigma$ field and the expression is evaluated around $T\simeq m_{\rm \sigma}$.

\begin{figure}
\begin{center}
\includegraphics[width = 0.6\textwidth]{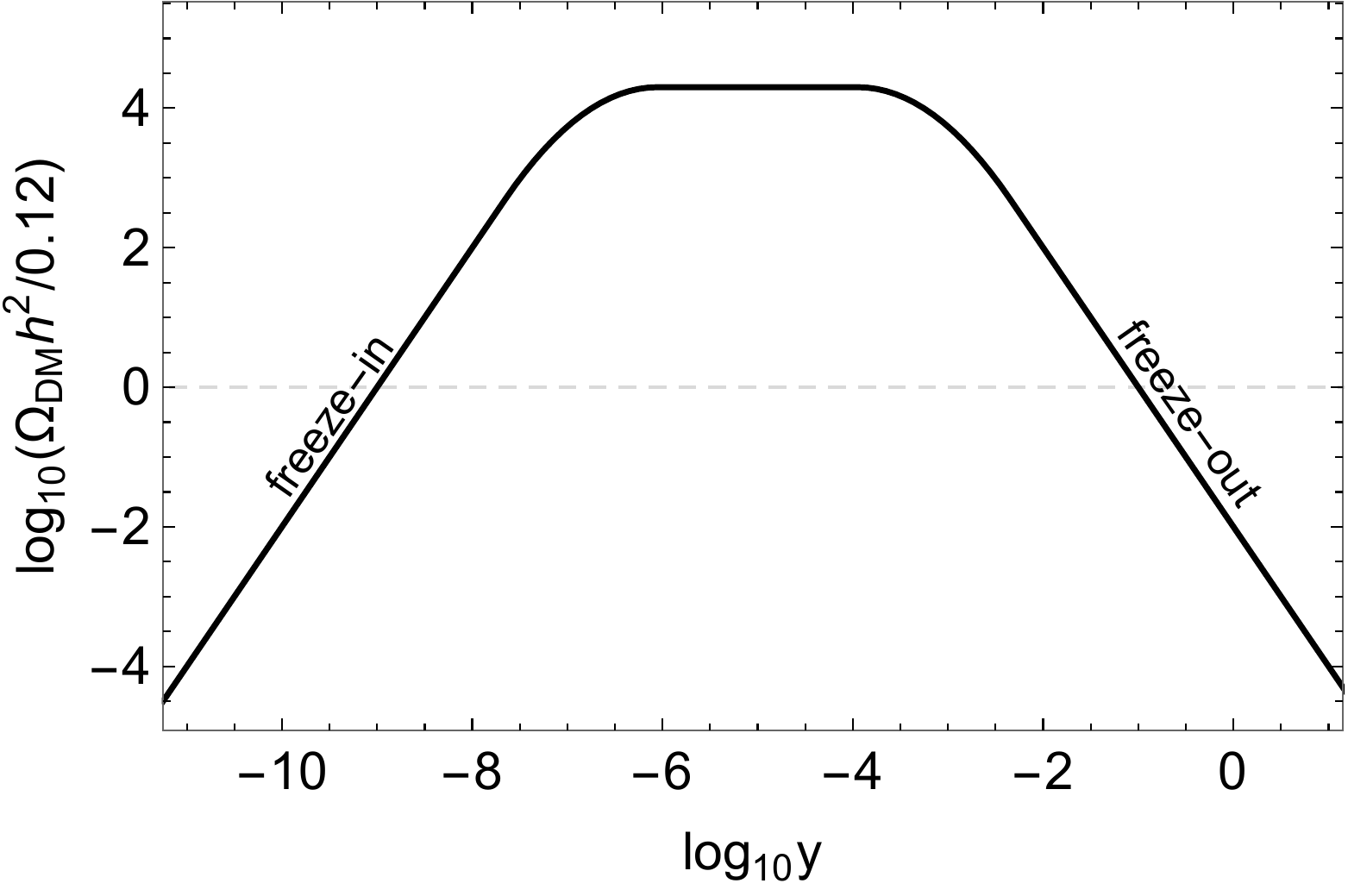}
\caption{A schematic representation of the DM relic density as a function of the coupling between DM and the visible sector. The plot corresponds to a transition between regimes 1 and 2 in Fig.~\ref{fig:phases}.}
\label{fig:volc}
\end{center}
\end{figure}

Taking then, for example, $g_{*s}\simeq g_*$ and $\Gamma_{\sigma\to \chi\chi}\simeq y^2\,m_{\rm \sigma}/(8\pi)$, where $y$ is the coupling strength between $\chi$ and $\sigma$, Eq.~\eqref{freezeinabundance} yields a parametric estimate for the coupling sufficient to produce a sizable DM abundance
\begin{equation}
\label{lowTestimate}
y\simeq 10^{-12}
\left(\frac{\Omega_{\rm \chi}h^2}{0.12}\right)^{1/2}\left(\frac{g_*}{100} \right)^{3/4}\left(\frac{m_\sigma}{m_{\rm \chi}}\right)^{1/2} .
\end{equation}
The implied small coupling value is compatible with the key assumption of the freeze-in scenario that the DM particles have not thermalized with the bath particles above $T\gtrsim m_{\rm \sigma}$. From Eq.~\eqref{lowTestimate} one can also see that the effect for the freeze-in yield in increasing the interaction rate between the visible sector and DM particles is opposite to that of the freeze-out scenario, where larger interaction rate implies smaller final abundance. This is illustrated in Fig.~\ref{FIvsFO}, and is further emphasized in Fig.~\ref{fig:volc}, where a schematic representation of the DM relic density as a function of the coupling $y$ between DM and the visible sector is shown. It should be noted that Eq.~\eqref{lowTestimate} gives an estimate of the magnitude of the coupling $y$ in the case where the initial DM abundance was zero or negligibly small. If this was not the case, Eq.~\eqref{lowTestimate} should be taken as an upper bound for $y$.

As another example of freeze-in, consider a scenario where the observed abundance is produced by $2\to 2$ annihilations. If the FIMP is a scalar and interacts with three scalar bath particles $\sigma_1$, $\sigma_2$, $\sigma_3$ via the operator $\zeta\,\chi\,\sigma_1\,\sigma_2\,\sigma_3$, the final abundance can be shown to be~\cite{Hall:2009bx}
\begin{equation}
\label{freezeinabundancescattering}
\Omega_{\rm \chi}h^2 \simeq 1.01\times10^{24}\frac{\zeta^2}{g_{*s}\sqrt{g_*}}\,,
\end{equation}
where the expression is evaluated at $T\simeq m_\chi$. In deriving Eq.~\eqref{freezeinabundancescattering} we assumed that the masses of $\sigma$ particles are negligible compared to the mass of $\chi$ and that the initial number density of $\chi$ was also negligible. To again generate a sizable DM abundance requires
\begin{equation}
\zeta\simeq 10^{-11}
\left(\frac{\Omega_{\rm \chi}h^2}{0.12}\right)^{1/2}\left(\frac{g_*}{100} \right)^{3/4}.
\end{equation}

The required coupling values and assumptions of the initial abundance are not the only differences between the freeze-in and freeze-out scenarios, as also the relation between the relevant mass scale and the bath temperature at the time of DM production is different. In the freeze-out mechanism the relic abundance is produced at $m_\chi/T \simeq 10\dots 30$, whereas for the freeze-in mechanism it arises during the epoch $m/T \simeq 2\dots 5$~\cite{Hall:2009bx}, where $m$ is the relevant mass scale in the DM production process. In the case of $\sigma\rightarrow \chi\chi$ decays this is $m_\sigma$, and for annihilations $\sigma\sigma\rightarrow \chi\chi$ it is max$(m_\chi, m_\sigma)$. Despite the fact that the decays and annihilations of visible sector particles can start early and gradually build up the DM abundance, the standard freeze-in involving only renormalizable operators is almost entirely an IR process. This can be seen, in the simplest case where the DM is produced by decays of bath particles, by either straightforwardly integrating Eq.~\eqref{FI_boltzmannY} or considering the estimate presented in Ref.~\cite{Petraki:2007gq}, where in order to find the comoving DM number density at $T\simeq m_\sigma$, one multiplies the number density of $\sigma$ particles by its decay rate and the time available for these decays to populate the DM abundance,
\begin{equation}
\frac{n_\chi}{T^3} \simeq t\,\Gamma_{\sigma\to \chi\chi} \simeq y^2\frac{M_{\rm P}}{m_\sigma} ,
\end{equation}
where $t\sim M_{\rm P}/T^2$ is the time-temperature relation for a radiation dominated Universe. The result shows that the freeze-in is essentially an IR process, and is indeed consistent with Eq.~\eqref{freezeinabundance}. The effect of annihilations at higher temperatures has been further discussed in Refs.~\cite{McDonald:2001vt, Yaguna:2011qn, Chu:2011be, Ayazi:2015jij, Drewes:2015eoa}.

The above calculation assumed that the initial number density of DM particles is negligible. Because the DM particles are assumed to have not been in thermal equilibrium with the visible sector particles in the early Universe, their production mechanism can be sensitive to the effect of non-renormalizable operators, the so-called \textit{ultraviolet freeze-in}~\cite{Choi:2005vq,Hall:2009bx, Yaguna:2011ei,Krauss:2013wfa,Elahi:2014fsa, Roland:2014vba, Roland:2015yoa, McDonald:2015ljz}, or to the initial conditions set by cosmic inflation~\cite{Enqvist:2014zqa, Dev:2014tla, Nurmi:2015ema, Kainulainen:2016vzv}. This is again in contrast to the freeze-out mechanism, where thermal equilibrium destroys all dependence on initial conditions. This important feature of the freeze-in mechanism can be used to constrain different models of feebly interacting DM, as we will discuss in Section~\ref{observations}. 

Also similarities between the freeze-out and freeze-in arise. Most importantly, they can both explain the observed DM abundance in a simple and yet predictive way. The simple form of the solutions~\eqref{freezeinabundance} and ~\eqref{freezeinabundancescattering} and the justifiable assumptions made in their derivation show that freeze-in is a perfectly adequate mechanism for generating the observed DM abundance. Furthermore, as with frozen-out particles, in the freeze-in scenario the frozen-in particle(s) may not be the particle(s) which comprise the final DM abundance: they may decay to actual DM particles, for example to sterile neutrinos, at a later stage~\cite{Shaposhnikov:2006xi, Kusenko:2006rh, Petraki:2007gq, Kang:2010ha, Merle:2013wta, Klasen:2013ypa, Adulpravitchai:2014xna, Merle:2014xpa, Merle:2015oja, Heikinheimo:2016yds, Konig:2016dzg}, as recently reviewed also in Refs.~\cite{Shakya:2015xnx,Adhikari:2016bei}. This option has recently become increasingly popular, as the Dodelson-Widrow (DW) mechanism~\cite{Dodelson:1993je} has now been ruled out as the production process for sterile neutrinos accounting for all of the observed DM abundance~\cite{Shakya:2015xnx, Adhikari:2016bei}. Also, the DM abundance can be produced by an \textit{asymmetric freeze-in}~\cite{Hall:2010jx, Hook:2011tk, Unwin:2014poa, Adshead:2016xxj, Hardy:2017wkr}, or it can consist of multiple components~\cite{DuttaBanik:2016jzv, Kainulainen:2016vzv, Heikinheimo:2016yds}. The frozen-in particles can also be related to the inflationary dynamics; the possibility of a FIMP driving inflation was recently studied in Ref.~\cite{Tenkanen:2016twd}. A dark sector not being in thermal equilibrium with the visible sector may also accommodate complicated dynamics of its own~\cite{Chu:2011be, Hochberg:2014dra, Bernal:2015ova, Bernal:2015xba, Heikinheimo:2016yds, Bernal:2017mqb}. This results in modifications in both the DM number density and its distribution function, with possibly interesting consequences for both direct and indirect detection, collider signatures, and cosmic structure formation. We will discuss these aspects in more detail in the following sections.

\subsection{Dark Freeze-out} 
\label{dark_FO}

In the first studies of the freeze-in mechanism, the effect of interactions in the dark sector was neglected when calculating the DM abundance~\cite{McDonald:2001vt, Hall:2009bx, Yaguna:2011qn, Blennow:2013jba, Campbell:2015fra, Kang:2015aqa}. However, it has later been shown to be important for both determining the initial conditions for DM production~\cite{Enqvist:2014zqa, Nurmi:2015ema, Kainulainen:2016vzv} and the subsequent evolution of DM number density after the yield from the visible sector is completed~\cite{Chu:2011be, Bernal:2015ova, Bernal:2015xba, Heikinheimo:2016yds, Bernal:2017mqb, Garcia-Cely:2017qpx}. Here we concentrate on the evolution of DM abundance produced initially by the decays and annihilations of visible sector particles and discuss the effect of dark sector couplings on the initial conditions for production in Section~\ref{observations}.

If the interactions in the DM sector are large, the DM particles may thermalize among themselves and other dark sector particles. This can happen independently of their interactions with the visible sector and does not require or imply that the dark sector particles would thermalize with the visible sector particles. In case the DM particles reach thermal equilibrium within the dark sector, their final abundance may become determined not by the initial freeze-in but by a freeze-out mechanism operating within the dark sector; the so-called \textit{dark freeze-out}~\cite{Dolgov:1980uu, Carlson:1992fn, Chu:2011be, Bernal:2015ova, Bernal:2015xba, Heikinheimo:2016yds, Bernal:2017mqb, Dolgov:2017ujf}. If the number-changing interactions are fast, they will lead to chemical equilibrium within the dark sector. This can happen even if the dark sector consists of only one particle species, i.e. the DM itself. The number-changing processes reduce the average momentum (temperature) of DM particles and increase their number density until equilibrium is reached. Both the DM abundance and its momentum distribution function may therefore change even though the coupling between the visible and dark sector has been effectively shut off.

\begin{figure}
\begin{center}
\includegraphics[width=.54\textwidth]{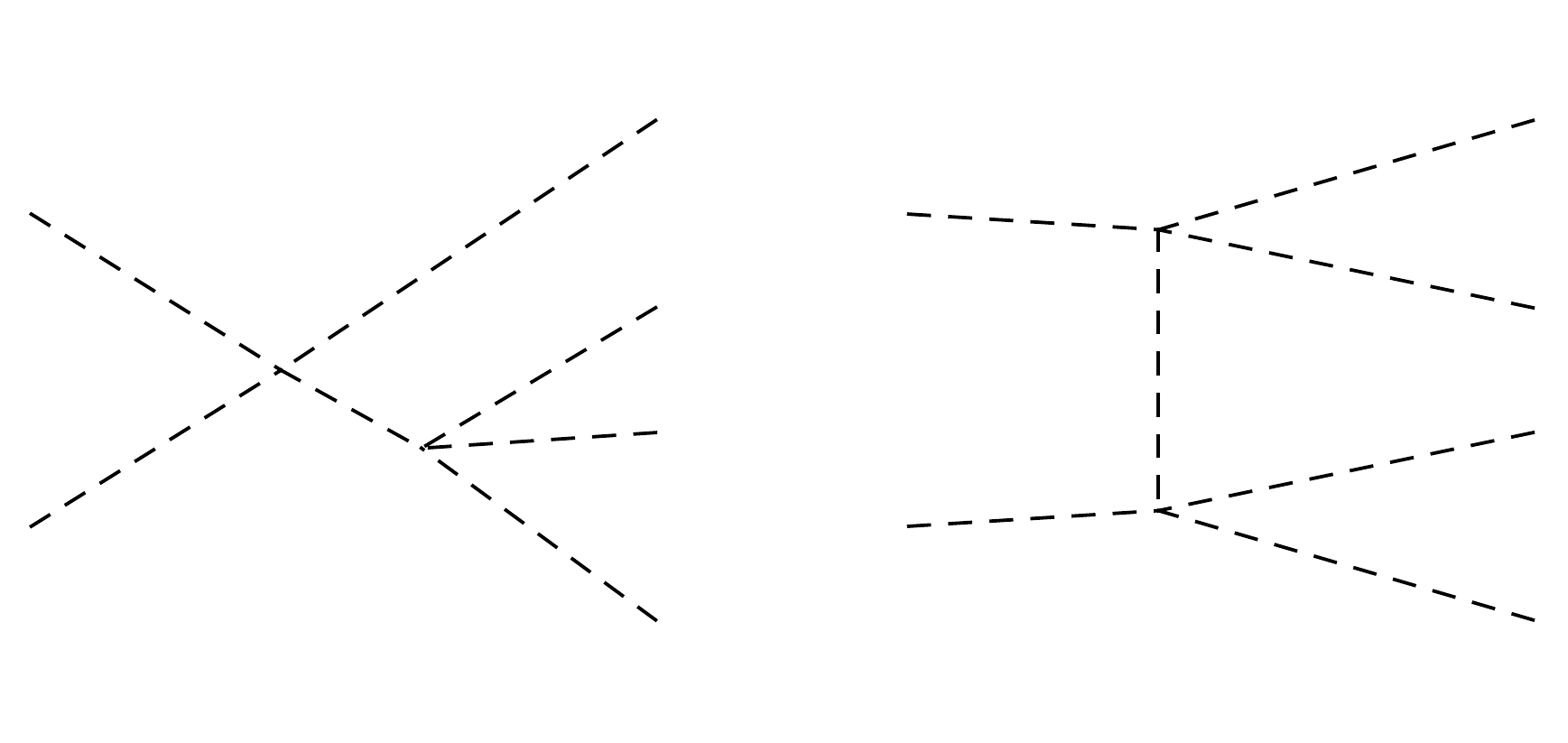}
\caption{Examples of Feynman diagrams for the $2\to 4$ scalar self-annihilation process.}
\label{diagram}
\end{center}
\end{figure}

For example, in case of DM self-interactions of the type $\lambda\chi^4$, it can be shown that thermalization of the dark sector within itself through $2\to4$ scatterings\footnote{For small $y$ the number density of DM particles at the time the initial freeze-in production ends is smaller than the corresponding equilibrium number density. Thus chemical equilibrium within the dark sector is reached by increasing the DM number density.} (See Fig.~\ref{diagram}) takes place if the self-coupling exceeds a critical value 
$\lambda_{\rm{crit.}}\equiv\lambda_{\rm{crit.}}(y,\,m_{\rm \sigma},\,m_{\rm \chi})$~\cite{Carlson:1992fn, Chu:2011be, Heikinheimo:2016yds}, where the parametric dependence assumes that the $\chi$ particles were produced by $\sigma \rightarrow \chi\chi$. For $\lambda<\lambda_{\rm{crit.}}$, the usual freeze-in picture is sufficient. If the self-coupling is larger than the critical value, the dark sector enters into thermal equilibrium with temperature $T_{\rm D}$, which in general differs from temperature $T$ of the baryon-photon fluid. In that period, the $2\to 4$ and the $4\to 2$ processes occur at the same rate.
The $2\to 4$ scatterings shut off when the DM particles become non-relativistic but the $4\to 2$ annihilations maintain the number density in its equilibrium value in a process called cannibalization~\cite{Carlson:1992fn}. During the cannibalization era, the DM particles increase their average momentum while their number density decreases. This increase of the momentum counteracts the effect of the redshift caused by the expansion of the Universe, resulting in a logarithmic scaling of the dark sector temperature as a function of the scale factor. The cannibalization era ends when the $4\to 2$ annihilation rate drops below the Hubble rate and the number density freezes out. As was first discussed in Ref.~\cite{Carlson:1992fn}, the freeze-out temperature of the cannibal phase is determined by the conservation of entropy in the dark sector, since in this case there are no other particle species available into which the DM could deposit its entropy as it becomes non-relativistic. 

\begin{figure}
\begin{center}
\includegraphics[width = 0.7\textwidth]{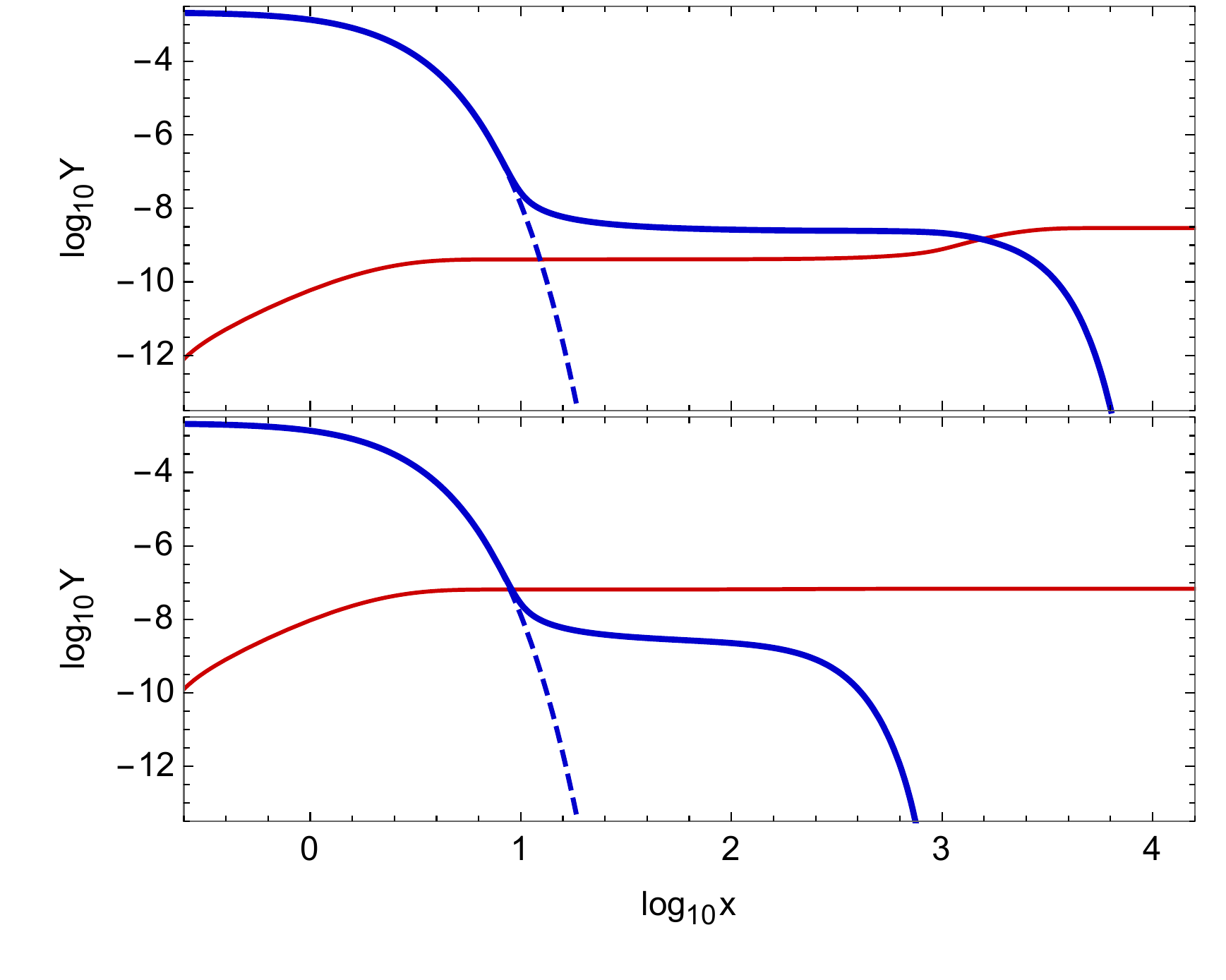}
\caption{Two examples of DM evolution due to dark sector dynamics: The red thin curve is the DM abundance and the blue thick curve is the comoving number density of a particle which reaches thermal equilibrium within the dark sector. Upper panel: If the coupling between the DM particle and the particle in equilibrium is very small, the particle production is dominated by the dark freeze-out and eventual decay of the particles in equilibrium. In this case freeze-in operating within the dark sector (the dark freeze-in) provides only a subdominant contribution to the final DM abundance. Lower panel: For larger coupling values -- but small enough for the DM particles not to thermalize with the particles in equilibrium -- the final DM abundance is given by the dark freeze-in. The figure is from Ref.~\cite{Heikinheimo:2016yds}.
}
\label{fig:boltzmann}
\end{center}
\end{figure}

The above picture relies on the conservation of entropy within the single DM species, but depending on the matter content of the dark sector, more intricate dynamics may arise. If the dark sector consists not only of one initially frozen-in particle, but has a more complicated structure, transfer of entropy between the various species becomes possible. The final DM abundance then depends on the hierarchy of coupling and mass values, both between the frozen-in particle and the visible sector, and within the dark sector. For example, a frozen-in dark sector particle which decays into actual DM particles may or may not undergo thermalization and subsequent dark freeze-out before the final DM abundance is produced; these two scenarios are illustrated in Fig.~\ref{fig:boltzmann}.
There, two examples of DM evolution due to a particular dark sector dynamics are shown: the red thin curve is the DM abundance and the blue thick curve is the comoving number density of a particle which reaches thermal equilibrium within the dark sector. In the upper panel it is shown the case where the coupling between the DM particle and the particle in equilibrium is very small; the particle production is dominated by the dark freeze-out and eventual decay of the particles in equilibrium. In this case freeze-in operating within the dark sector (the \textit{dark freeze-in}) provides only a subdominant contribution to the final DM abundance.
However, in the lower panel it is shown the case corresponding to larger coupling values (but small enough for the DM particles not to thermalize with the particles in equilibrium), where the final DM abundance is given by the dark freeze-in.
These aspects have been investigated further in Refs.~\cite{Chu:2011be, Williams:2012pz, Bernal:2015ova, Bernal:2015xba, Heikinheimo:2016yds, Bernal:2017mqb}.

\subsection{Reannihilation} 
\label{reannihilation}
\begin{figure}
\begin{center}
\includegraphics[width = 0.7\textwidth]{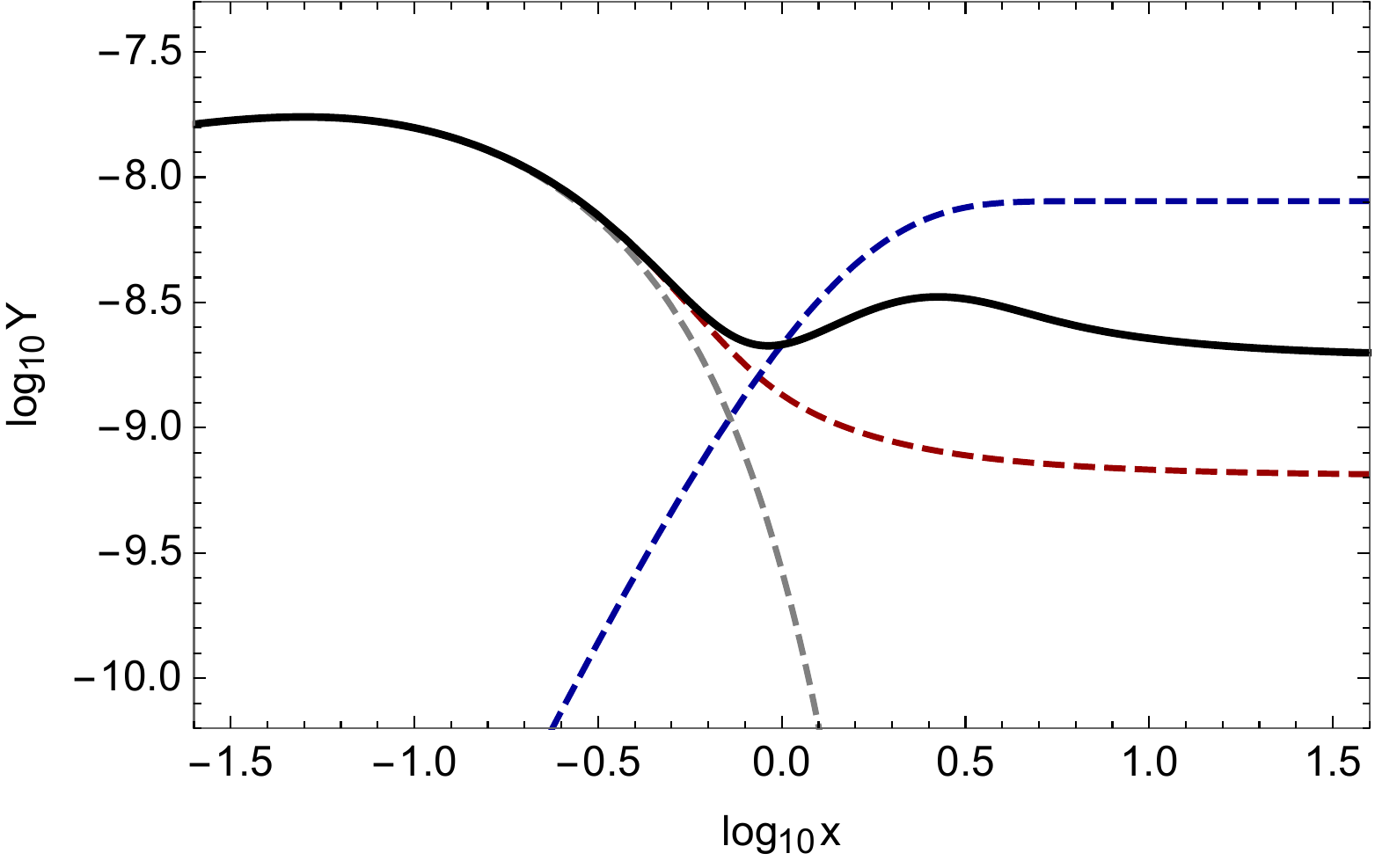}
\caption{Example of the evolution of DM abundance in the case of reannihilation. Shown are the actual abundance (black solid line), equilibrium abundance (gray dashed line), frozen-out abundance (red dashed line), and frozen-in abundance (blue dashed line).}
\label{fig:reannihilation}
\end{center}
\end{figure}

The final possibility we consider here is the one where DM abundance changes due to a process dubbed \textit{reannihilation}~\cite{Cheung:2010gj, Cheung:2010gk, Chu:2011be, Bernal:2015ova}. This is a scenario where the dark sector thermalizes within itself but the dark freeze-out would occur already before the yield from the visible sector has ended. In that case, the on-going particle yield keeps increasing the DM relic abundance and thus resumes annihilations of dark sector particles. The DM number density first follows the equilibrium density and then the so-called \textit{quasi-static equilibrium} density, until it intercepts a critical value $Y_{\rm crit}\simeq \langle v\sigma\rangle\,s\,Y_{\rm eq}^2/H$, (see e.g. Ref.~\cite{Chu:2011be}), and finally freezes out. This is illustrated in Fig.~\ref{fig:reannihilation}.

The reannihilation mechanism was introduced in Ref.~\cite{Cheung:2010gj} and has then been studied in various contexts including supersymmetry~\cite{Cheung:2010gk}, Higgs portal~\cite{Chu:2011be,Bernal:2015ova}, kinetic mixing portal~\cite{Chu:2011be}, and hidden vector DM models~\cite{Bernal:2015ova}. Further details on how the process depends on properties of DM and the possible mediator between the dark and visible sectors can be found from the above references.

\section{Models for Freeze-in}
\label{models}
Many beyond the SM scenarios feature candidates for particle DM. Results from precision electroweak tests and flavor physics provide 
stringent constraints for models where the new degrees of freedom feel the 
SM gauge interactions. To alleviate these constraints, one turns to model 
building utilizing new states which are neutral under the SM 
gauge interactions. The new degrees of freedom can thermalize with the SM 
fields provided that suitable mediators between the visible and dark
sectors exist. On the other hand, if the coupling is feeble enough, the dark 
sector remains out of equilibrium or may thermalize among itself if strong
enough particle number-changing self-interactions exist, as discussed in Section~\ref{DMgenesis}. 

In this Section we will briefly review some general frameworks for frozen-in DM. We will start with a short overview of portal models where the coupling 
between a hidden singlet sector and the visible SM sector arises from the 
interactions with the Higgs scalar or from the mixing between a new hidden 
$U(1)$ gauge boson and the SM hypercharge. The DM particle itself in these cases 
can be a scalar, a fermion or a vector. Then, we will discuss in more detail the specific possibility that the DM 
particle is a singlet fermion, i.e. a sterile neutrino, possibly with a small mixing 
with the active neutrino
species. We will also briefly consider other model building paradigms where the 
freeze-in mechanism has been applied to explain the observed DM abundance. 

\subsection{Portal Models}

\subsubsection{Higgs Portal}

A simple paradigm for hidden sectors minimally coupled with the SM is a singlet scalar interacting with the SM only via the Higgs field~\cite{Silveira:1985rk, McDonald:1993ex, Burgess:2000yq}, i.e. via the Higgs portal. Models of this type can be extended by considering the DM candidate to be a singlet vector or a fermion. The interaction between the DM and the visible sector is via the Higgs field $\Phi$ and can, in the simplest form, be parametrized by one of the following operators:
\begin{eqnarray}
  \frac{\lambda_{\rm{hs}}}{2}|\Phi|^2s^2,\quad \frac{\lambda_{\rm{h\psi}}}{\Lambda}|\Phi|^2\bar\psi\psi\quad\text{and}\quad
  \frac{\lambda_{\rm hA}}{2}|\Phi|^2A_\mu A^\mu\,.
\end{eqnarray}
In addition to the above portal coupling the assumed mass of the DM particle will enter the cross sections for the relevant interaction rates. While these effective operators provide a simple parametrization of DM interactions, there are several immediate shortcomings.

First, in the above effective theory formulation an ad hoc discrete symmetry for scalar DM candidate must be postulated in order to guarantee its stability. Second, the Lagrangian for the singlet fermion is non-renormalizable and requires a UV completion. Third, while the vector Lagrangian looks renormalizable, it is really not the case as the vector mass must be put in by hand. This in turn leads to undesirable high energy behavior and violation of unitarity. Also this model can be UV completed to resolve these problems, and due to larger particle content, any UV completion will imply very different phenomenology than the simple effective theory form. More complete models also allow one to address further problems, like the order of the electroweak phase transition, in addition to the existence of DM~\cite{Cline:2012hg,Alanne:2013dra, Alanne:2014bra}.

As a concrete example of freeze-in production of DM in such a model, let us consider a real scalar singlet $s$ which is odd under a $\mathbb{Z}_2$ symmetry while all the SM particles are even. The most general renormalizable scalar potential reads
\begin{equation}
\label{scalarpot}
  V(\Phi,s)=\mu_{\rm{h}}^2|\Phi|^2+\lambda_{\rm{h}}(\Phi^\dagger \Phi)^2+\frac{\mu_{\rm{s}}^2}{2}s^2
  +\frac{\lambda_{\rm{s}}}{4}s^4+\frac{\lambda_{\rm{hs}}}{2}|\Phi|^2s^2 ,
\end{equation}
where $H$ is the usual SM Higgs doublet with the standard kinetic terms. If the portal coupling is very weak, $\lambda_{\rm hs}\lesssim 10^{-7}$, the singlet sector does not equilibrate with the SM. Furthermore, if $2m_{\rm s}\leq m_{\rm h}$, the abundance of $s$ particles is produced via freeze-in through the Higgs boson decay channel $h\rightarrow ss$. The final $s$ abundance is
\begin{equation}
\label{scalarDMyield}
  \frac{\Omega_{\rm{s}}h^2}{0.12}\simeq 5.3\times 10^{21}\,\lambda_{\rm{hs}}^2
  \,\left(\frac{m_{\rm{s}}}{{\rm{GeV}}}\right) ,
\end{equation}
in analogy with Eq. \eqref{freezeinabundance}. If the mass hierarchy is reversed, $m_{\rm s}\gg m_{\rm h}$, the result is given by an expression similar to Eq. \eqref{freezeinabundancescattering}. However, as discussed in Section~\ref{dark_FO}, the whole picture can change considerably if the $2\to 4$ scalar self-interactions are active, i.e. if $\lambda_{\rm s}$ was sufficiently large. In that case, the $s$ particles may thermalize within itself after the initial DM production from the SM sector has ended and the DM number density can change even after the coupling between DM and the SM has been effectively shut off. That being the case, the final DM abundance is determined not only by the freeze-in, but also by the dark freeze-out occurring in the hidden sector~\cite{Chu:2011be, Bernal:2015ova, Bernal:2015xba, Heikinheimo:2016yds, Bernal:2017mqb}.

From Eq.~\eqref{scalarDMyield} one can see that obtaining the correct DM abundance generically requires a very small coupling, $\lambda_{\rm hs}\lesssim 10^{-9}$. One might worry that the smallness of the portal coupling poses a naturalness problem and makes this scenario unappealing for model building. However, as the running of $\lambda_{\rm hs}$ is usually negligible (see e.g. Ref.~\cite{Kainulainen:2016vzv}), the requirement of having $\lambda_{\rm hs}\ll 1$ at the scale where DM production takes place does not impose a fine-tuning problem.  Furthermore, it has been argued that the small coupling connecting a nearly isolated hidden sector to the SM can be seen as technically natural due to the enhanced space-time symmetry group in the limit of decoupling~\cite{Foot:2013hna}.

As an extension of the simplest framework, Eq.~\eqref{scalarpot}, consider the case where the hidden sector consists of a real singlet pseudoscalar field $s$ and a fermion $\psi$, invariant under the parity transformation $s(t,\,x)\to -s(t,\,-x)$ and $\psi(t,\,x)\to \gamma^0\psi(t,\,-x)$~\cite{Kainulainen:2016vzv}. The fermionic part of the portal sector is
\begin{equation}
  {\cal L}_\psi=\bar\psi\,(i\,\gamma^\mu\partial_\mu-m_\psi)\,\psi+i\,g\,s\,\bar\psi\,\gamma_5\,\psi\,,
  \label{pseudo}
\end{equation}
and we again assume the form~\eqref{scalarpot} for the scalar potential. In principle, the fermion could be interpreted as a special case of a  sterile neutrino. Here we will take this particular model just as a simple illustrative  case and consider sterile neutrino DM in more detail in the next Subsection. If the singlet scalar is light enough, $2\,m_{\rm s}<m_{\rm h}$, the final yield becomes
\begin{equation}
  \frac{\Omega_{\rm{DM}}h^2}{0.12}=5.3\times 10^{21}\,\lambda_{\rm{hs}}^2\,N\,
  \left(\frac{m_{\rm{DM}}}{{\rm{GeV}}}\right)\,.
\end{equation}
If DM is composed of singlet scalar, $m_{\rm{s}}<2\,m_\psi$, then $m_{\rm{DM}}=m_{\rm{s}}$ and $N=1$.
On the other hand, if DM is the fermion,
$m_{\rm{s}}>2\,m_\psi$, then $m_{\rm{DM}}=m_\psi$ and $N=2$ accounts for the fact that
in each subsequent $s$ decay two fermions $\psi$ are produced. However, this result can again change if the number-changing interactions in the hidden sector are sizable~\cite{Heikinheimo:2016yds}.

Similarly, one can consider Higgs portal vector DM~\cite{Hambye:2008bq, Hambye:2009fg, Arina:2009uq, Bernal:2015ova}. To have a well defined UV theory, one assumes that the vector is a gauge boson of a hidden gauge symmetry which in the simplest case is $SU(2)$. This is an appealing choice, since then there is a residual $SO(3)$ symmetry guaranteeing the stability of the vector DM. Furthermore, in comparison to the $U(1)$ gauge symmetry there is no mixing with the SM hypercharge which would need to be tuned to be small to agree with experiments~\cite{An:2014twa, An:2013yfc, Redondo:2013lna}, as we will discuss below.

One interesting subset of portal models consists of scale invariant theories. Such models have been applied to explain dynamically the origin of the electroweak scale and connect the electroweak and DM scales, see e.g. Refs.~\cite{Bardeen:1995kv, Meissner:2006zh, Chang:2007ki, Hambye:2007vf, Foot:2007iy, AlexanderNunneley:2010nw, Hur:2011sv, Englert:2013gz, Heikinheimo:2013fta, Gabrielli:2013hma, Carone:2013wla, Hambye:2013sna, Foot:2013hna,  DiChiara:2015bua,Karam:2015jta,Karam:2016rsz, Marzola:2017jzl} and references therein. They have also attracted some further attention within the context of producing the DM abundance via the freeze-in mechanism~\cite{McDonald:2001vt, Kang:2014cia, Kang:2015aqa, Heikinheimo:2017ofk}. One attractive feature of this approach is that as the number of free parameters is reduced, the predictability of the model increases. Let us briefly review this model building paradigm with an example of a self-interacting, scale invariant hidden sector consisting of a real singlet $\mathbb{Z}_2$-symmetric scalar.

The singlet scalar is coupled to the SM via the usual Higgs portal, Eq.~\eqref{scalarpot}. For the scalar self-interaction, one assumes $\lambda_{\rm{s}}>0$ for stability and for the portal coupling $\lambda_{\rm{hs}}>0$ in order not to induce a vev for the singlet scalar $s$. The coupling between the dark and visible sectors is assumed small, $\lambda_{\rm{hs}}\lesssim 10^{-7}$. Scale generation in the dark sector is driven by the spontaneous symmetry breaking in the Higgs sector, which we here parametrize with the vacuum expectation value $v$ of the Higgs field.\footnote{In a more complete setting within the classically scale invariant framework, also this scale should be dynamically generated, and various realizations can be found in the references above.} As $\sqrt{2}\,\Phi=(0,v+h)$, where $v=246$~GeV, the singlet gains a mass
\begin{equation}
  m_{\rm{s}}=\sqrt{\frac{\lambda_{\rm{hs}}}{2}}\,v.
\end{equation}

The DM abundance, produced via the freeze-in mechanism, is in this case
\begin{equation}
  \frac{\Omega_{\rm{s}}h^2}{0.12}\simeq 5\times 10^{19}\,\lambda_{\rm{hs}}^2
  \left(\frac{m_{\rm{s}}}{10~{\rm{MeV}}}\right)\simeq 6\times
  10^{23}\,\lambda_{\rm{hs}}^{5/2}\,.
\end{equation}
Requiring the abundance to match the observed relic density implies $\lambda_{\rm s}\simeq 3\times 10^{-10}$ and $m_{\rm s}\simeq 2$ MeV. Finally, the strength of singlet scalar self-interactions is
\begin{equation}
  \frac{\sigma_{\rm{s}}}{m_{\rm{s}}}=\frac{9\lambda_{\rm{s}}^2}{32\pi \,m_{\rm{s}}^3}\simeq 0.1
  \left(\frac{\lambda_{\rm{hs}}}{10^{-9}}\right)^{-3/2}\left(\frac{\lambda_{\rm{s}}}
  {0.1}\right)^2\frac{{\rm{cm}}^2}{{\rm{g}}}.
\end{equation}
Taking for example, $\sigma_{\rm{s}}/m_{\rm{s}}=1$ cm$^2$/g (as discussed in Section \ref{DMSI}) we see that this scenario uniquely
fixes all model parameters: $\lambda_{\rm{s}}\simeq 0.13$, $\lambda_{\rm{hs}}\simeq 3\times 10^{-10}$ and $m_{\rm{s}}\simeq 2$ MeV. Because of this feature, the
scenario has been called the \textit{FIMP miracle}~\cite{McDonald:2001vt, Kang:2014cia, Kang:2015aqa}.

However, the situation is more complicated than originally studied in Refs.~\cite{McDonald:2001vt, Kang:2014cia, Kang:2015aqa} if the DM self-interactions are large. In that case the hidden sector may again thermalize within itself and the DM number density can change after the yield from the visible sector has completed. This changes the requirements for the model parameters and renders the FIMP miracle to a \textit{WIMP miracle of the second kind}. This was originally pointed out in Ref.~\cite{Bernal:2015xba} and studied in detail in Ref.~\cite{Heikinheimo:2017ofk}.

\subsubsection{Kinetic Mixing Portal}

Another portal paradigm arises if the hidden sector consists of a new $U(1)$ gauge field, a dark photon. Then the Lagrangian will include a gauge invariant term
\begin{equation}
\label{darkphoton}
  {\cal L}\supset \epsilon\,F^{\mu\nu}_Y {F_D}_{\mu\nu},
\end{equation}
where $F^{\mu\nu}_Y$ and $F^{\mu\nu}_D$ are the field strengths of the SM hypercharge and the new dark $U(1)$ interaction, respectively, and the mixing parameter $\epsilon \ll 1$. 
The dark photon is assumed to have mass, $m_{\rm{A}}$, induced in a gauge invariant way by the St\"uckelberg mechanism~\cite{Kors:2005uz} or by a dark Higgs. If the dark photon is light, $m_{\rm{A}}<2\,m_{\rm{e}}$, as required by stability, then the DM abundance can be produced by the freeze-in mechanism through photon scattering with electrons and positrons or through electron-positron annihilation~\cite{Pospelov:2008jk, Redondo:2008ec}.
In the relevant mass range, the required magnitude of the coupling $\epsilon$ is already in the ballpark tested by direct detection searches~\cite{An:2014twa}.

On the other hand, such feebly interacting vector DM abundance can also be produced by quantum fluctuations during inflation~\cite{Graham:2015rva},
\begin{equation}
\frac{\Omega_{\rm{A}}h^2}{0.12}\simeq \left(\frac{m_{\rm{A}}}{1\,{\rm{keV}}}\right)^{1/2}
\left(\frac{H_\ast}{10^{12}\,{\rm{GeV}}}\right),
\end{equation}
where $H_\ast$ denotes the inflationary scale. In this case the vector is produced with a power spectrum peaked at intermediate wavelengths. This suppresses the small wavelength isocurvature perturbations below the observed level, while at long wavelengths the perturbations are the usual nearly scale-invariant adiabatic perturbations of the inflaton field. This is in stark contrast to the scalar case~\cite{Nurmi:2015ema, Kainulainen:2016vzv, Heikinheimo:2016yds}, as we will discuss further in Section~\ref{cosmological_signatures}.

The setting~\eqref{darkphoton} can be also enlarged by adding matter fields (either scalar or fermion) charged under the hidden $U(1)$ and singlet under the SM interactions. These new matter fields will provide new DM candidates. The freeze-in production of the DM abundance in such setting has been investigated in Refs.~\cite{Chu:2011be, Berger:2016vxi, Essig:2015cda, Fradette:2014sza,Chu:2013jja}. 

\subsubsection{Other Portal Models}

If the dark sector has scalar and fermion fields, it is possible to have portals via right-handed SM leptons 
as in e.g. Ref.~\cite{Bai:2014osa}. A further realization of such leptonic portal arises in 
the so-called scotogenic model (or radiative seesaw model) \cite{Ma:2006km} and its variants~\cite{Molinaro:2014lfa}.
This model is particularly appealing as it connects DM with the neutrino phenomenology. We will review models of this type in more detail in the next Subsection.

Before concluding this Subsection, we note that hidden sectors, coupled with the SM via a scalar portal, could be also composite~\cite{Cline:2013zca, Soni:2016gzf, Acharya:2017szw}, with UV description in terms of gauge and fermion fields singlet under all the SM charges.\footnote{It is also possible that hidden sector contains millicharged fields, and this possibility also results in viable DM candidates~\cite{Kouvaris:2013gya}.} 
In general there are many reasons which make composite DM an
appealing alternative: First, stability of DM can arise automatically
from global symmetries in the underlying theory. Second, strongly coupled
theories are natural candidates for self-interacting DM required to
address issues observed in small scale structure formation. Third, the DM scale arises similarly as the confinement scale in QCD and is therefore
natural. Fourth, the rich spectrum expected to arise in any strongly coupled
composite theory may provide unique observational cues. As an alternative and much studied model building paradigm hidden sectors also appear in the context of supersymmetric theories~\cite{DeSimone:2010tr}.

\subsection{Sterile Neutrinos}

Sterile neutrinos are the most studied DM candidates in the context of freeze-in mechanism. They are a well-motivated extension of the SM, because they may not only comprise the observed DM abundance but also accommodate a mechanism for generating Majorana masses for the active SM neutrinos, for example by the seesaw mechanism. For a comprehensive list of references on sterile neutrinos, we refer to a recent review~\cite{Adhikari:2016bei}. Our discussion on sterile neutrinos as frozen-in DM will also overlap with Ref.~\cite{Shakya:2015xnx}. 
However, we will provide updates by taking the newer literature into account, and include discussion on typical initial conditions and number-changing interactions.

In the context of freeze-in production of DM, consider the Lagrangian
\begin{equation}
\label{sterileneutrinoL}
\mathcal{L} \supset y\,\bar{L}\,\tilde{\Phi}^{\dagger}\,\nu_{\rm R} + m\,\bar{\nu}_{\rm R}^c\,\nu_{\rm R} + s\,\bar{\nu}_{\rm R}^c\,(y_S + i\,y_P)\,\nu_{\rm R} + {\rm h.c.} + V(\Phi,s)\,,
\end{equation}
where $L=(e,\nu_{\rm L})^T$ and $\tilde{\Phi}=\epsilon \Phi^*$, where $\epsilon$ is the antisymmetric $SU(2)$ invariant tensor and $\Phi$ is again the SM Higgs doublet. The first term in Eq.~\eqref{sterileneutrinoL} gives rise to a Dirac after the electroweak symmetry breaking, the second term is a Majorana mass and the third term describes both a scalar and pseudo-scalar interactions between a singlet scalar $s$ and a sterile neutrino $\nu_{\rm R}$. The last term is the scalar potential, which we leave unspecified. Here both $s$ and $\nu_{\rm R}$ are assumed to be sterile, i.e. singlets under the SM gauge symmetries.\footnote{Another possibility is to introduce a new electromagnetically charged scalar, so that the freeze-in production of sterile neutrinos occurs via decay of the charged scalar to a sterile neutrino and a charged SM lepton~\cite{Frigerio:2014ifa}. A related scenario, where an unstable FIMP decays into sterile neutrino DM in a similar model, was studied in Ref.~\cite{Okada:2015kkj}.}

There are essentially two ways of producing sterile neutrino DM in this model setup: via the weak interaction if the sterile neutrinos mix with the active neutrinos, or by decays or (semi-)annihilations of the scalar $s$. Although the first mechanism does not actually fulfill our definition of freeze-in (non-thermal DM density produced by decays and annihilations of particles in thermal equilibrium), it is worth discussing here as a property of a generic sterile neutrino model. The main production mechanism in the case of a sizable mixing angle $\theta$ is the DW mechanism, which non-resonantly generates the sterile neutrino abundance around $T\simeq 150$ MeV~\cite{Dodelson:1993je}. This means that sterile neutrino properties can be constrained both by indirect detection experiments and their free-streaming length, i.e. observations of small scale structure formation. Indeed, measurements of the Lyman-$\alpha$ forest have been shown to rule out $m_{\nu_{\rm R}} \lesssim 16$~keV if the DW-produced sterile neutrinos account for all DM~\cite{Baur:2015jsy}. On the other hand, the active-sterile mixing allows for $\nu_{\rm R}$ to decay into an active neutrino and a photon, and observations of the X-ray sky have been shown to rule out $m_{\nu_{\rm R}}\gtrsim 4$ keV if all DM consists of DW-produced sterile neutrinos \cite{Boyarsky:2006ag}. It is therefore evident that sterile neutrinos produced by the DW mechanism cannot account for all DM, although it has been claimed that they can still comprise $\mathcal{O}(10)\%$ of the observed abundance~\cite{Boyarsky:2008xj, Boyarsky:2008mt}.

In contrast to the DW mechanism, the freeze-in mechanism where the DM abundance is produced by decays and (semi-)annihilations of the singlet scalar allows for the mixing angle to be arbitrarily small. This both alleviates the Lyman-$\alpha$ and X-ray constraints\footnote{Also other ways to resurrect the sterile neutrino as a viable DM candidate accounting for all of the observed abundance exist. They have been discussed exhaustively in Refs.~\cite{Shakya:2015xnx, Adhikari:2016bei}.} and makes the sterile neutrinos stable in the limit $\theta\to 0$. As a finite mixing angle allows for decays of $\nu_{\rm R}$ into the SM particles, the DW mechanism can at the same time produce sterile neutrinos whose lifetime is longer than the age of the Universe and which account for all DM only when their mass is not larger than $m_{\nu_{\rm R}} \sim \mathcal{O}(1)$ keV. Introducing a new scalar $s$ and considering the freeze-in as a production mechanism for sterile neutrinos then allows them to have basically any mass, which opens up the parameter space and new possibilities for model building.

Introducing the scalar as a mediator also allows for the possibility that the typical sterile neutrino momenta are much smaller than in the DW scenario, as they can now be produced at earlier times. This has immediate consequences for cosmic structure formation and leads to less severe constraints on sterile neutrino properties. Also, if freeze-in is responsible for producing most of sterile neutrino DM but one also allows for a small non-zero mixing angle, the scenario can lead to indirect astrophysical signatures, as we will discuss in more detail in Section~\ref{direct_indirect_detection}.

For these reasons frozen-in sterile neutrinos have been studied in a number of papers during the recent years. Early studies on the topic include Refs.~\cite{Shaposhnikov:2006xi, Bezrukov:2008ut, Bezrukov:2009yw} and~\cite{Kusenko:2006rh, Petraki:2007gq}, where the authors considered decays and annihilations of the inflaton field or a generic scalar $s$ as a production mechanism for sterile neutrinos, respectively. More recent studies have considered similar scenarios where the scalar $s$ either decays into sterile neutrinos while in equilibrium~\cite{Klasen:2013ypa, Molinaro:2014lfa, Abada:2014zra, Alanne:2014bra, Roland:2014vba, Kang:2014mea, Arcadi:2014dca, Humbert:2015epa,Merle:2015oja, Matsui:2015maa, Adulpravitchai:2015mna, Biswas:2016bfo, Konig:2016dzg, Roland:2016gli, Biswas:2016iyh, Biswas:2017tce}, or where it first undergoes a freeze-in itself~\cite{Merle:2013wta, Abada:2014zra, Merle:2014xpa, Adulpravitchai:2014xna, Roland:2014vba, Merle:2015oja, Matsui:2015maa, Heikinheimo:2016yds, Konig:2016dzg, Roland:2016gli, Biswas:2016iyh, Heikinheimo:2017ofk}. Recently, sterile neutrinos undergoing or related to freeze-in have also been studied in supersymmetric models~\cite{Kang:2010ha, Roland:2015yoa, Shakya:2016oxf, Roland:2016gli}. A useful estimate for the DM abundance in these scenarios can be given as~\cite{Kusenko:2006rh, Petraki:2007gq}
\begin{equation}
\label{sterileNabundance}
\frac{\Omega_{\nu_{\rm R}}h^2}{0.12} \simeq \left(\frac{y}{10^{-8}} \right)^3\frac{\langle s\rangle}{m_{\rm s}} ,
\end{equation}
when the scalar with mass $m_{\rm s}$ decays into sterile neutrinos while in equilibrium, and as~\cite{Merle:2015oja}
\begin{equation}
\label{sterileNabundance2}
\frac{\Omega_{\nu_{\rm R}}h^2}{0.12} \simeq \frac{30}{g_*(T_{\rm prod})}\left(\frac{\lambda_{\rm hs}}{10^{-12}}\right)^2
\end{equation}
when the scalar itself first undergoes freeze-in. Here $\lambda_{\rm hs}$ is again the coupling between the singlet scalar and the Higgs doublet, as in Eq.~\eqref{scalarpot}.

However, similarly to what was discussed in the previous Section, production of frozen-in sterile neutrinos can be sensitive to physics at higher energy scales. In Refs.~\cite{Kainulainen:2016vzv, Heikinheimo:2016yds} it was shown that cosmic inflation sets generic initial conditions for sterile neutrino production also in the case where $s$ is not the inflaton but an energetically subdominant spectator field. Constraints on the model parameter space following from the non-observation of primordial DM isocurvature in the CMB were derived in the same works. Furthermore, as discussed in Section~\ref{dark_FO}, if the dark sector exhibits sufficiently large (self-)interactions, the sterile neutrino abundance produced from the SM sector energy density may subsequently change due to number-changing interactions such as $ss\to \nu_{\rm R}\bar{\nu}_{\rm R}$, invalidating the estimates of Eqs.~\eqref{sterileNabundance} and~\eqref{sterileNabundance2} and requiring a more detailed dark freeze-out calculation as described in Section~\ref{dark_FO}. These aspects were recently studied in Refs.~\cite{Heikinheimo:2016yds, Heikinheimo:2017ofk}, where it was shown that also in this case sterile neutrinos can comprise all DM.

Even though frozen-in sterile neutrinos have been shown to easily account for all DM and can in certain cases lead to observable signatures (see Section~\ref{direct_indirect_detection}), their effect on cosmic structure formation is currently not clear. The determination of DM momentum distribution function is important not only for astrophysics in ascertaining the exact effect that DM has on cosmic structure formation, but also because such structure formation may be the only way to test models where DM interacts only feebly with the SM sector.  Regarding this, an important step forward was taken in Refs.~\cite{Merle:2014xpa, Merle:2015oja, Merle:2015vzu, McDonald:2015ljz, Biswas:2016iyh, Konig:2016dzg}, although they neglected the effects of thermalization within the dark sector, namely the subsequent evolution of DM number density due to number-changing interactions as discussed above in Section~\ref{dark_FO}, or thermal corrections to particle decay rates as discussed in Ref.~\cite{Drewes:2015eoa}. However, in Ref.~\cite{Merle:2015oja} an important fact was pointed out: frozen-in sterile neutrinos cannot, in general, be simply defined as `hot', `warm', or `cold' DM, as in reality the average momentum as a measure of the free-streaming length can be too na\"{i}ve an estimate for determining the effect that frozen-in sterile neutrinos have on cosmic structure formation. Determining the origin and properties of sterile neutrino DM and their implications on structure formation are currently under active study.

\subsection{Other Models}

Finally, we will briefly review other models considered in the literature. In the minimal supersymmetric SM model the neutralino as the lightest supersymmetric particle is a well-studied WIMP candidate. However, as the WIMP scenario has become very constrained, also in supersymmetric models the freeze-in mechanism has been considered as the production mechanism for DM abundance. These include studies on neutralino and axino DM~\cite{Covi:2002vw,Cheung:2011mg, Bae:2014rfa, Medina:2014bga, Co:2015pka}, goldstino DM~\cite{Monteux:2015qqa}, gravitino DM~\cite{Cheung:2011nn, Hall:2012zp, Hall:2013uga, Fan:2014gxa, Co:2016fln, Benakli:2017whb,Dudas:2017rpa,Argurio:2017joe} and photino DM~\cite{Kolda:2014ppa}. Also frozen-in decaying DM~\cite{Kang:2010ha} and asymmetric freeze-in production of DM generating also the baryon asymmetry in the Universe~\cite{Hall:2010jx} have been considered in a supersymmetric framework.

Other studies on asymmetric freeze-in include Refs.~\cite{Hook:2011tk, Unwin:2014poa, Hardy:2017wkr}. In these scenarios the dark and visible sectors are in thermal equilibrium with themselves but not with each other. The out-of-equilibrium condition necessary for baryogenesis is provided by the temperature difference between the two sectors and the baryon and DM number asymmetries are then produced by CP violating decays or annihilations from one sector to another. In Refs.~\cite{Choi:2005vq,Hall:2009bx,Yaguna:2011ei,Krauss:2013wfa,Elahi:2014fsa, Roland:2014vba, Roland:2015yoa, McDonald:2015ljz} freeze-in via non-renormalizable operators was considered, while in Ref.~\cite{Mambrini:2013iaa} an alternative scenario where the mass of the portal scalar is larger than the reheating temperature was proposed. In these scenarios the DM abundance is produced during the transition from inflaton domination to radiation domination. Also, Ref.~\cite{Dedes:2017shn} considered freeze-in production of radiative light DM.

In Ref.~\cite{Garcia-Cely:2017oco} it was shown that freeze-in is a viable production mechanism for a Majoron DM particle, which is a pseudo-Nambu-Goldstone boson resulting from spontaneous breaking of a global $U(1)_{B-L}$ symmetry. Similar pseudo-Nambu-Goldstone boson DM production via freeze-in was studied in Ref.~\cite{Frigerio:2011in}. 
Freeze-in in the context of $U(1)_{B-L}$ and $U(1)_{L_\mu-L_\tau}$ symmetry was also considered in Refs.~\cite{Kaneta:2016vkq, Biswas:2017tce} and~\cite{Biswas:2016yjr}, respectively. In Refs.~\cite{Kaneta:2016vkq,Heurtier:2016iac} frozen-in DM particle is a right-handed neutrino, whereas in Ref.~\cite{Aoki:2015nza} DM is produced non-thermally via decays of right-handed neutrinos. Furthermore, Refs.~\cite{Babichev:2016hir, Babichev:2016bxi} studied a bimetric theory, which contains in addition to the massless graviton a second spin-2 particle which is massive. It was shown that the new massive spin-2 particle can be DM and its abundance can be produced via freeze-in.
Finally, in Refs.~\cite{Garny:2015sjg, Tang:2016vch, Babichev:2016hir, Babichev:2016bxi} it was shown that gravitational DM can be produced by the exchange of gravitons via the ultraviolet freeze-in mechanism.

\section{Observational Properties of Frozen-in Dark Matter}
\label{observations}

The defining property of frozen-in DM is the weakness of the coupling between the SM and the DM particle. By construction, the coupling must be so feeble that the DM particle never reaches thermal equilibrium with the visible sector. On the first look it therefore appears that such models must be exceedingly difficult to experimentally observe by means other than gravity. However, as we shall discuss below, even these feebly coupled models may be subjected to experimental scrutiny, and features of frozen-in dark sectors can be uncovered via terrestrial and astrophysical experiments.

\subsection{Direct and Indirect Detection Signatures}
\label{direct_indirect_detection}
FIMP DM scenarios are characterized by tiny couplings with the SM particles.
They are therefore very challenging to test at colliders or fixed-target experiments, but novel low-threshold direct detection techniques have a particularly important opportunity to probe them in the near future.
In fact, light DM has to have a large number density in order to match the measured DM abundance, enhancing therefore the detection rates. Multiple experimental setups have recently been suggested for the detection of elastic and inelastic scatterings of DM in the mass range from keV to MeV~\cite{Essig:2011nj, Essig:2012yx, Lee:2015qva, Essig:2015cda, Green:2017ybv, Emken:2017erx, Essig:2017kqs, Graham:2012su, Hochberg:2016sqx, Hochberg:2015pha, Hochberg:2015fth, Hochberg:2016ajh, Derenzo:2016fse, Schutz:2016tid, Knapen:2016cue, Essig:2016crl, Kouvaris:2016afs, McCabe:2017rln}.
In particular, the typical DM-electron cross sections for an MeV FIMP DM could be tested by some next generation experiments~\cite{Essig:2011nj, Alexander:2016aln, Bernal:2017mqb, Essig:2017kqs}.
Finally, due to the possibility of absorbing a dark photon, the direct detection constraints for light vector DM in the $U(1)$ vector portal model extend down to values of the kinetic mixing coupling constant $\epsilon$ relevant for freeze-in~\cite{An:2014twa}.

Indirect detection signals can result from decay or annihilation processes of DM particles. The feeble couplings characteristic for freeze-in models can result in very long lifetimes, so that decaying DM can be naturally embedded in the freeze-in paradigm. This idea has been pursued in the context of frozen-in heavy neutrino DM, where energetic SM neutrinos could be observed in the IceCube as DM decay products~\cite{Fong:2014bsa,Fiorentin:2016avj, Chianese:2016smc}. Decaying DM interpretation of a spectral feature at $E\simeq 3.55$~keV observed in X-ray observations from several DM dominated sources~\cite{Bulbul:2014sua,Boyarsky:2014jta} has been studied in the context of frozen-in DM in Refs.~\cite{Queiroz:2014yna, Baek:2014poa, Farzan:2014foo, Arcadi:2014dca, Merle:2014xpa, Roland:2015yoa, Kang:2014cia}.
In general, X-ray and gamma-ray observatories provide a powerful and independent probe of light DM, in the ballpark relevant for FIMP DM~\cite{Essig:2013goa}.

Due to the smallness of the couplings between the DM and the SM particles in the typical freeze-in models, the annihilation signal from DM DM $\rightarrow$ SM SM processes is usually very suppressed and not observable. However, it is possible that annihilation processes take place within a frozen-in dark sector, where stable DM particles $\chi$ annihilate into unstable mediator particles $\phi$, which in turn decay into the SM. In this case, the cross section of the process $\chi\chi\rightarrow \phi\phi$ does not have to be very much suppressed, while the dark sector remains secluded from the SM due to small couplings between the mediator $\phi$ and the SM. Such mechanism has been applied in the context of the galactic center gamma ray excess in Ref.~\cite{Heikinheimo:2014xza}. A model that is able to fit both the X-ray line and the gamma-ray excess with a frozen-in DM sector was presented in Ref.~\cite{Biswas:2015bca}

\subsection{Collider Signatures}

There are intensive searches for WIMP DM at colliders. Typical analyses follow model-dependent scenarios based on full or simplified models, as well as more model-independent searches for a mono-X accompanied by missing energy, as discussed in Section~\ref{sec:collsearches}. These experiments are typically not sensitive to freeze-in models, due to the small production cross section implied by the small couplings inherent in this framework.

However, in some cases the freeze-in scenario could result in visible collider signals. A unique signature of a feebly coupled dark sector is the appearance of particles with macroscopic lifetimes~\cite{Brooijmans:2012yi, Arcadi:2013aba, Arcadi:2014tsa, Hessler:2016kwm, Ghosh:2017vhe}. These can be searched for in collider experiments with designated searches for the so-called displaced signatures: tracks appearing away from the collision axis, long-lived particles decaying in the calorimeter, or disappearing tracks.
Searches for long-lived particles have been presented both by ATLAS~\cite{Aad:2012pra, Aaboud:2016uth} and CMS~\cite{Chatrchyan:2013oca, Khachatryan:2016sfv}.
These effects have been considered in a scenario where the freeze-in mechanism operates during an early matter-dominated era after inflation~\cite{Co:2015pka, Evans:2016zau}. In this case the initial DM abundance produced via the freeze-in mechanism is diluted due to entropy production at the onset of the following radiation dominated era, resulting in a stronger than usual coupling between the dark sector and the SM in order to yield the correct DM abundance.

In many cases, the frozen-in DM is coupled to the SM via a mediator particle, which itself is not necessarily feebly coupled to the visible sector. This mediator particle could then produce observable signatures in collider experiments. These signatures have been considered in Refs.~\cite{Arcadi:2013aba, Molinaro:2014lfa, Arcadi:2014dca, Arcadi:2014tsa, Ayazi:2015jij, Hessler:2016kwm}.

For lower mass DM, the strongest constraints follow from limits on anomalous scattering at proton and electron fixed-target experiments such as LSND~\cite{deNiverville:2011it, Kahn:2014sra, Auerbach:2001wg} and E137~\cite{Batell:2014mga, Bjorken:1988as},
and on proton beam dump measurements with MiniBooNE~\cite{Aguilar-Arevalo:2017mqx}.
For specific mass ranges, limits on invisible pion~\cite{Atiya:1992sm}, kaon~\cite{Pospelov:2008zw, Artamonov:2009sz} and $J/\Psi$~\cite{Ablikim:2007ek} decays are also significant.

\subsection{Astrophysical Signatures}

As discussed in Section~\ref{DMSI}, the observed small scale structure of the Universe may not be correctly reproduced within the collisionless cold DM paradigm. A potential solution to the small scale problems is given by warm DM, where structure formation is suppressed below the free-streaming scale of the light DM particles in the early Universe. However, warm DM is severely constrained by the Lyman-$\alpha$ forest data~\cite{Irsic:2017ixq}. This discrepancy could be alleviated if the DM velocity distribution is not thermal, which is possible if DM is produced via freeze-in. Such possibility has been explored in Refs.~\cite{Merle:2014xpa, Merle:2015oja, Merle:2015vzu, McDonald:2015ljz, Biswas:2016iyh, Konig:2016dzg}. Another way to suppress the formation of structures at small scales is provided by late kinetic decoupling of DM~\cite{Bringmann:2006mu, Bringmann:2009vf, Aarssen:2012fx, Bringmann:2016ilk}. In this scenario the DM particles are kept in kinetic equilibrium with relativistic species, either from the visible or the dark sector, until a sufficiently late time, leading to similar effects as the warm DM free streaming. An attempt to capture all relevant particle physics input for the formation of small scale structure has recently been carried out in the effective theory of structure formation (ETHOS) framework~\cite{Cyr-Racine:2015ihg, Vogelsberger:2015gpr}.

The models for warm DM typically contain light DM particles, with $m_\text{DM}\sim\mathcal{O}(\rm keV)$. Fermion DM below this mass scale is constrained by considerations of the phase space density in small halos, the so called Tremaine-Gunn bound: $m_{\rm DM}\gtrsim 0.5$~keV~\cite{Tremaine:1979we, Boyarsky:2008ju}. A slightly less constraining bound has been obtained in a recent analysis, where assumptions of the scaling of the DM halo radius with respect to the stellar component were relaxed~\cite{DiPaolo:2017geq}.

Self-interacting DM produced via the freeze-in mechanism has been considered in Refs.~\cite{McDonald:2001vt, Campbell:2015fra, Kang:2015aqa, Bernal:2015ova, Bernal:2015xba, Heikinheimo:2016yds, Bernal:2017mqb,Heikinheimo:2017ofk}. In this scenario it is important to account for the number-changing interactions within the dark sector, which play an important role in determining the DM abundance as discussed above in Section~\ref{dark_FO}. Observations of the small scale structure and cluster mergers may provide important insight into the non-gravitational interactions inside the dark sector, allowing us to probe the properties of DM regardless of the weakness of the coupling between the dark and visible sectors.

The production and emission of feebly interacting light particles can also be constrained by observations of certain astrophysical objects like stars and supernovae.
For DM particles lighter than $\sim 100$~keV, energy losses from stars dominate the constraints~\cite{An:2014twa, An:2013yfc, Redondo:2013lna}, whereas MeV-scale DM is also constrained by supernovae; for the latest bound see Refs.~\cite{Chang:2016ntp, Hardy:2016kme}.

\subsection{Cosmological Signatures}
\label{cosmological_signatures}

The defining feature of the freeze-in scenario is that the DM never enters thermal equilibrium with the SM heat bath. An important consequence is that the dark sector remains sensitive to primordial initial conditions, as opposed to the WIMP scenario, where thermal equilibrium erases any information about the possible non-thermal initial state of the system. This leads to unique observational constraints on the dark sector, that are not present for models where DM is a frozen-out thermal relic.

First, if the inflaton field couples directly to the DM particles, reheating of the dark sector may result in an overabundance of DM. This feature can be used to set limits on the strength of the inflaton-DM coupling, as discussed in Refs.~\cite{Kofman:1994rk, Dev:2013yza, Bezrukov:2008ut, Shaposhnikov:2006xi, Adshead:2016xxj}. For a scenario where a frozen-in DM particle is the inflaton, see Ref.~\cite{Tenkanen:2016twd}.

Second, even in the absence of any direct coupling between the DM and inflaton fields, the dark sector fields may be light spectator fields during inflation, and they will generally obtain non-vanishing primordial field values in different Hubble patches. In the absence of DM thermalization with the visible sector, any such primordial perturbations in the DM density spectrum are not washed out but remain physically observable, leaving an imprint in the CMB. These isocurvature fluctuation modes are a generic feature of models where feebly coupled scalar fields exist in the dark sector~\cite{Nurmi:2015ema, Kainulainen:2016vzv, Heikinheimo:2016yds}. They arise from density fluctuations of dark sector scalar fields, in the case that the energy density stored in a scalar field at the end of inflation is deposited into DM. These modes are severely constrained by the CMB power spectrum~\cite{Ade:2015lrj}.

\begin{figure}
\begin{center}
\includegraphics[width=.58\textwidth]{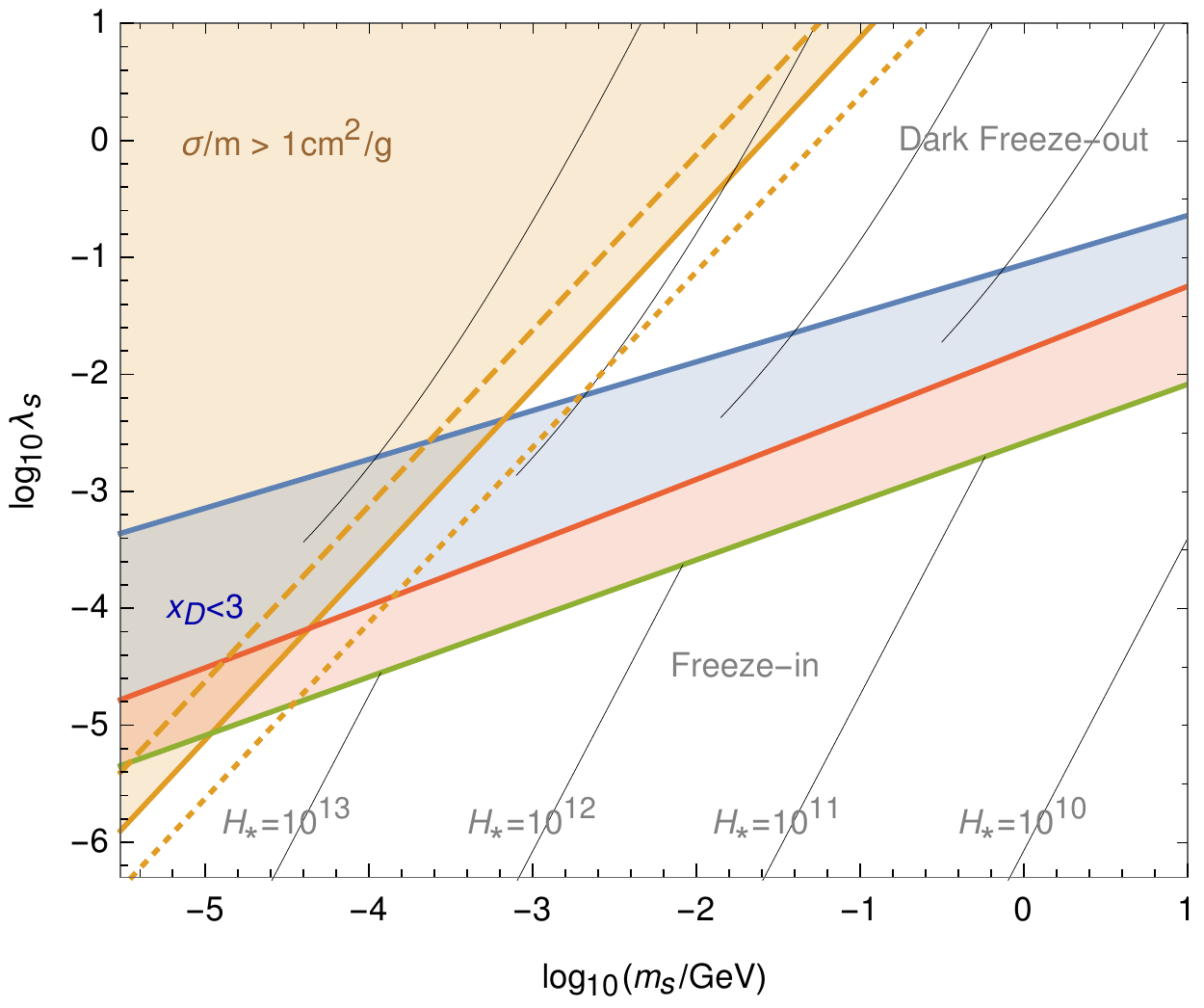}
\caption{The self-interaction bound and isocurvature constraints for the scalar DM scenario. The self-interaction limit, $\sigma_{\rm DM}/m_{\rm DM} < 1$~cm$^2$/g, is shown by the yellow shaded region in the top left corner together with the $\sigma_{\rm DM}/m_{\rm DM} = 10$, $0.1$~cm$^2$/g contours (dashed and dotted, respectively), and the isocurvature constraints by the black contours for $H_* = 10^{13}$, $10^{12}$, $10^{11}$, $10^{10}$~GeV from left to right. The region to the right of the isocurvature contours is ruled out by the CMB constraint, for the given values of the inflationary scale $H_*$. For details, see Ref.~\cite{Heikinheimo:2016yds}.}
\label{HP_constraints}
\end{center}
\end{figure}

The amplitude of the isocurvature signal is controlled by the flatness of the dark sector scalar potential. Therefore the upper bound on the isocurvature perturbations can be expressed as a lower limit on the scalar self-coupling~\cite{Kainulainen:2016vzv,Heikinheimo:2016yds}:
\begin{equation}
\label{isocurvature_ia_bound}
\lambda_{\rm s} \gtrsim 0.008\left(\frac{m_{\rm DM}}{\rm GeV}\right)^{8/3}\left(\frac{H_*}{10^{11}~{\rm GeV}}\right)^4 ,
\end{equation}
in the scalar potential of Eq.~\eqref{scalarpot}. If the scalar itself is the DM particle, this imposes a lower limit on the DM self-interaction strength, as depicted in Fig.~\ref{HP_constraints}. However, these limits depend on the currently poorly constrained value of the Hubble rate at the end of inflation, $H_*$, resulting in significant uncertainty on the constraints that can be drawn from these considerations. Future probes of the primordial tensor perturbation spectrum will be important for probing the scale of inflation, and therefore also constraining freeze-in models via the isocurvature limit.

Due to the feeble couplings generically present in the freeze-in scenario, long lived particles are a rather common feature in these models. If they decay into the SM, the energy deposited into the SM photon bath between BBN and photon decoupling eras can lead to observable effects in the light element abundances or in the CMB~\cite{Poulin:2016anj}. These effects have been considered in the context of freeze-in models in Ref.~\cite{Berger:2016vxi}. Constraints from BBN and CMB for a frozen-in millicharged DM have been considered in Refs.~\cite{Fradette:2014sza,Gabrielli:2015hua}.

Finally, the recent experimental success in observing gravitational waves has opened up a new avenue for observational cosmology, which will be further enhanced by, for example, the upcoming LISA mission~\cite{Audley:2017drz}. Gravitational waves produced in dark sector phase transitions could be observed with the LISA experiment, regardless of the weakness (or even absence) of non-gravitational interactions between the dark and visible sectors~\cite{Schwaller:2015tja,Caprini:2015zlo}, thus providing an additional observational handle on an otherwise secluded dark sector, such that appears in models of freeze-in.
\section{Summary}
\label{conclusions}

Despite the decades spanning program to search 
for DM, its interactions with ordinary matter, beyond gravity, remain unknown. The lack of a confirmed experimental detection signal is a crucial challenge to the WIMP DM paradigm. However, the FIMP scenario, where DM is a very weakly interacting particle produced by the freeze-in mechanism in the early Universe, remains as a perfectly adequate framework for explaining what are the properties of DM and how was its observed abundance generated.

In this review we presented the freeze-in mechanism and its known variations, dark freeze-out and reannihilation, and gave a summary of research conducted in the field so far. We discussed various model frameworks, including portal models, sterile neutrinos, asymmetric and supersymmetric FIMPs. We also discussed the observational properties and general testability of very weakly interacting DM.

We stress that the mechanisms and models considered in this review are both well-motivated and viable candidates for explaining the observed DM abundance. As discussed above, seeking further for their signatures in particle physics experiments and both astrophysical and cosmological observations may lead to unexpected new probes in the future and finally reveal what DM is.

\section*{Acknowledgments}

We thank Xiaoyong Chu, Andreas Goudelis, Alexander Merle, Maximilian Totzauer and Stephen West for careful reading and valuable comments on the manuscript.
This project has received funding from the Academy of Finland grant 267842, the National Science Centre (Poland) research project, decision  DEC-2014/15/B/ST2/00108, the European Union's Horizon 2020 research and innovation programme under the Marie Skłodowska-Curie grant agreements 674896 and 690575; and from Universidad Antonio Nari$\rm{\tilde{n}}$o grant 2017239. N.B. is partially supported by the Spanish MINECO under Grant FPA2014-54459-P, T.T. is supported by the U.K. Science and Technology Facilities Council grant ST/J001546/1, and V.V. by the Estonian Research Council grant IUT23-6 and ERDF Centre of Excellence project No TK133.

\tiny
\bibliography{freezeinrev.bib}
\end{document}